\documentclass[11pt,a4paper]{article}
\pdfoutput=1 
\usepackage[margin=1in]{geometry}
\usepackage{mciteplus} 
\usepackage{graphicx,rotating,hyperref,slashed,amsmath,xcolor,amssymb,amsfonts,colortbl,cite}  

\hypersetup{colorlinks, bookmarksopen, bookmarksnumbered,
citecolor=verdes, linkcolor=bluc, pdfstartview=FitH, urlcolor=rossos}

\usepackage{amsmath,amssymb,bm}
\usepackage{graphicx}
\usepackage{booktabs}
\usepackage{hyperref}
\usepackage{xcolor}
\usepackage{enumitem}
\usepackage{caption}
\usepackage{physics}

\hypersetup{colorlinks,bookmarksopen,bookmarksnumbered,
linkcolor=blue,pdfstartview=FitH,urlcolor=blue,citecolor=blue}
\numberwithin{equation}{section}
\usepackage{mciteplus}

\def\be{\begin{equation}}
\def\ee{\end{equation}}

\def\bea{\begin{eqnarray}}
\def\eea{\end{eqnarray}}

\newcommand{\ii}{\mathrm{i}}
\newcommand{\Tcal}{\mathcal{T}}

\newcommand{\mpl}{M_{\rm Pl}}
\newcommand{\hinf}{H_{\rm inf}}
\newcommand{\ks}{k_*}
\newcommand{\kc}{\kappa_c}
\newcommand{\ms}{m_\sigma}
\newcommand{\sigbar}{\bar\sigma}
\newcommand{\PT}{\mathcal{T}_\sigma}
\newcommand{\Ps}{\mathcal{P}_S}

\begin{document}


\vspace{1truecm}
 \begin{center}
\boldmath
{\textbf{\LARGE  
{ 
Inflationary Kicks and Coulomb Filtering of Spectator Dark Matter  
}}
}
\unboldmath
\end{center}
\unboldmath

\vspace{-0.2cm}

\begin{center}
\vspace{0.1truecm}

\renewcommand{\thefootnote}{\fnsymbol{footnote}}
\begin{center} 
{\fontsize{13}{30}
\selectfont  
 Martina La Rosa$^{a\!}$, \, Rasheshwari Paul Chowdhury$^{b\!}$, \,Gianmassimo Tasinato$^{c,d\!}$ \footnote{\texttt{g.tasinato2208.at.gmail.com}}
} 
\end{center}


\begin{center}
\vskip 6pt
\textsl{$^a$
Dipartimento di Fisica e Astronomia `Galileo Galilei', Universit\`a di Padova, 35131 Padova, \\
     Istituto Nazionale di Fisica Nucleare, Sezione di Padova, 35131 Padova, Italy}
\\
\textsl{$^b$ Department of Physics and Electronics, CHRIST (Deemed to be University), Bengaluru, India}\\
\textsl{$^c$ Physics Department, Swansea University, SA2 8PP, UK}
\\
\textsl{$^{d}$ Dipartimento di Fisica e Astronomia, Universit\`a di Bologna,\\
 INFN, Sezione di Bologna,  viale B. Pichat 6/2, 40127 Bologna,   Italy}
\vskip 4pt
\end{center}
\hskip0.1cm
\end{center}

\begin{abstract}
\noindent
Light spectator fields provide a compelling inflationary origin for scalar dark matter, but their primordial fluctuations often generate excessive isocurvature perturbations on large scales. We develop a novel, exactly solvable filtering mechanism based on a tower of $\delta-$function contributions to the spectator mass, which we call kicks, whose dense limit converges to a smooth Coulomb-like profile. The  filter produces a  isocurvature spectrum during inflation that is strongly suppressed on large scales, rises with momenta as $k^3$ in the infrared, and, after a rapid turnover, smoothly approaches a constant plateau at small scales. 
This construction is  motivated by a discrete translational invariance, admits
an EFT regime in which radiative corrections can be systematically controlled, 
and  can find  a concrete realization in terms of an inflationary massive-clock model. 
We embed the filter in a minimal dark-matter scenario in which a spectator condensate oscillates about the minimum of a quadratic potential and behaves as pressureless matter. In this framework, the spectator dark-matter mass is directly related to the turnover scale of the inflationary isocurvature spectrum. We determine the resulting dark-matter abundance and analyze the relevant isocurvature and backreaction constraints. The most restrictive realization, in which the spectator condensate is generated during inflation, favors a relatively low inflationary scale.
\end{abstract}

\section{Introduction}
\label{sec:introduction}

The particle nature of dark matter and its connection with the physics of the
early Universe remain among the central open questions of cosmology.  A
particularly economical possibility is that dark matter originates from a
scalar field already present during inflation.  If sufficiently weakly coupled to the visible sector, such a field can remain energetically subdominant during
inflation and later survive as a coherent condensate, eventually oscillating
about the minimum of its potential and behaving as pressureless matter.  This
spectator-dark-matter picture is attractive because it directly links the relic
abundance to inflationary initial conditions, without requiring thermal
equilibrium with the Standard Model. See e.g. \cite{Gorbunov:2011zzc} for a textbook
discussion. 

The same simplicity, however, creates a well-known difficulty.  A light
spectator $\sigma$ field acquires nearly scale-invariant fluctuations of order
$H_{\rm inf}/(2\pi)$ during inflation.  If these fluctuations are uncorrelated
with the adiabatic mode, they subsequently source cold-dark-matter
isocurvature perturbations and are tightly constrained by the CMB
\cite{Planck:2018jri,Planck:2018vyg}.  A number of mechanisms have been proposed
to suppress or reshape spectator fluctuations, including self-interactions
that modify the inflationary spectator distribution
\cite{Markkanen:2018gcw,Padilla:2019fju}, effective masses generated by
couplings to the inflaton or to curvature \cite{Garcia:2023awt,Chakraborty:2025lyp}, spectra whose
power is predominantly concentrated on short scales
\cite{Alonso-Alvarez:2018irt}, and post-inflationary mechanisms that alter the
conversion between field and density fluctuations \cite{Batell:2026sml}.
More generally, a sufficiently blue isocurvature spectrum points toward
non-trivial time dependence in the spectator mass sector: for a linear spectator
with a constant mass throughout the relevant history, the largest measurable
blue tilt is  more restricted \cite{Chung:2015tha,Chung:2015pga}.

\smallskip

In this work we investigate a different possibility.
We build a scenario in which the spectator mass receives
a series of $\delta-$function contributions, which we dub pulses, or kicks, appropriately located along 
the inflationary time evolution. In the dense
limit, they converge to a Coulomb-type spectator mass. The resulting spectrum
for spectator fluctuations is exactly solvable: it exhibits a filtering mechanism
for large-scale fluctuations, whose amplitude is  suppressed with respect to the 
small-scale ones. Interesting, the resulting shape of the spectrum has the
correct profile to match -- and theoretically motivate -- phenomenological profiles
recently considered in the literature in related contexts. This mechanism can then be applied to build a minimal, and observationally consistent model of inflationary dark matter from a light spectator field.

\smallskip

\begin{figure}[t!]
 \centering
   \includegraphics[width=0.7\linewidth]{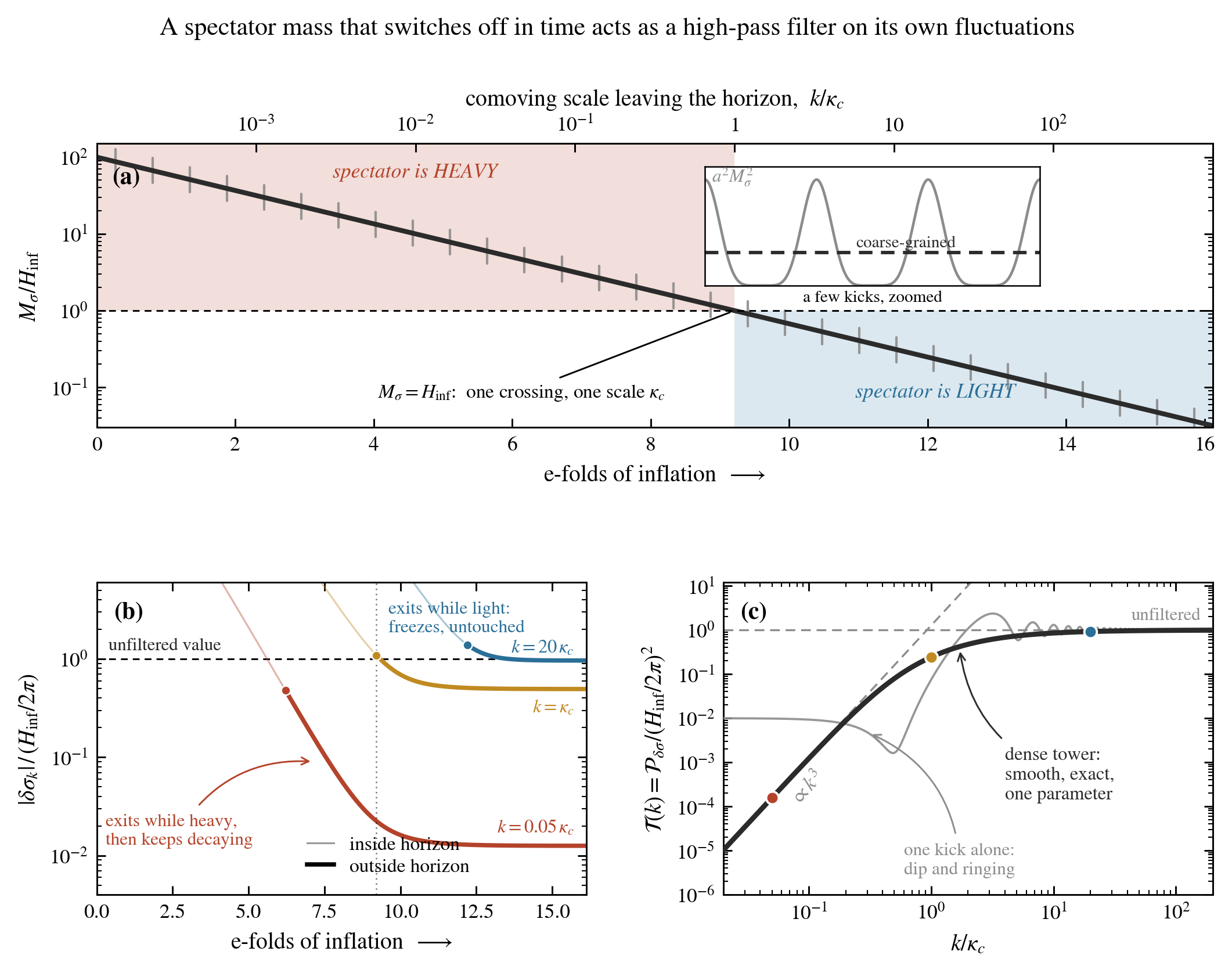}
 \caption{\small  The filtering mechanism. \textbf{(a)} A tower of brief mass
kicks, equally spaced in e-folds, coarse-grains to a mass envelope that falls
through $M_\sigma = H_{\rm inf}$  once, defining the  comoving
scale $\kappa_c$ discussed in the main text. \emph{Inset:} an individual kick and its average.
\textbf{(b)} The mechanism in three modes: each freezes on leaving the horizon
(dots), but a mode that exits while the spectator is heavy continues to decay
and ends up  suppressed, while a mode exiting after the field has
become light retains the standard amplitude $H_{\rm inf}/2\pi$. Curves are
obtained by  numerical integration from Bunch--Davies initial conditions.
\textbf{(c)} The resulting transfer function. A single kick leaves a dip and
oscillatory ringing (grey); the dense logarithmic tower reproduces the exact
Coulomb result (black), smooth and controlled by $\kappa_c$ alone.}
 \label{fig:single-positive-kick-summaryV1}
\end{figure}

Specifically,
we start our work motivating the case for  short, positive contributions to the
spectator mass during inflation, approximated
by a tower of $\delta-$function terms.  Such events may be generated by sharp
features, transient changes in a multifield trajectory, symmetry-breaking
transitions, repeated particle-production events, or other localized departures
from adiabatic evolution.  When their duration is short compared with the time
scale of the modes of interest, they can be represented as instantaneous mass
kicks.  The mode evolution then becomes a temporal-scattering problem, and an
arbitrary sequence of kicks is described by an ordered product of transfer matrices,
following a formalism well explored in the physics of scattering.

In fact, a single
 kick already illustrates the basic mechanism: it mixes the two
independent mode solutions (the growing and decaying mode) and can strongly suppress sufficiently long
wavelengths, while generically producing a dip followed by oscillatory ringing.
With two or more kicks, the state generated by one event is scattered by the
next, and the phases accumulated between successive events produce coherent
interference.  The finite-pulse problem is exactly tractable, but its spectrum
contains several scales and retains information about the detailed pulse
configuration. We then show that a much simpler structure emerges when the kicks form a dense logarithmic
tower. In the dense limit, we can coarse-grain the system: their contribution can be resummed to a Coulomb-type
time-dependent mass term for the spectator field. The corresponding spectrum
of spectator fluctuations can be exactly solved during inflation. Large-scale modes that cross
the horizon early when the spectator mass is large with respect to the Hubble
parameter get an amplitude suppression, and scale as $k^3$ for small values
of momenta. Small-scale modes that instead leave the horizon late during the inflationary phase have a mass much smaller than Hubble, and retain a scale-invariant spectrum. The smooth transition between the two momenta
regimes -- which depend on a single comoving  scale $\kc$ --  results exactly computable.   
We graphically summarize the filtering method in 
Fig.~\ref{fig:single-positive-kick-summaryV1}.

\smallskip

An important part of our analysis is to clarify the theoretical status of this
construction.  In the discrete description, the spectator  enjoys an exact
shift symmetry between successive events.  Non-derivative
radiative corrections are then tied to the same symmetry-breaking source
rather than generating an unrelated bulk potential.  Finite pulse widths,
peak masses, separations between events, physical momenta, and the EFT cutoff
provide explicit control parameters.  The logarithmic arrangement of the
pulses is not protected by the spectator shift symmetry itself and should instead
be inherited from the sector generating the source.  In the dense regime,
these considerations organize the renormalization of the effective Coulomb
profile and identify a parametric domain in which the low-energy description
is under perturbative control.
We  give an explicit proof-of-principle realization of the logarithmic
tower in terms of an additional heavy spectator acting as an inflationary
clock.  A field oscillating rapidly while adiabatically tracking a
time-dependent minimum naturally produces events equally spaced in physical
time and hence logarithmically spaced in conformal time.  With an appropriate
slowly varying envelope, its coupling to the spectator $\sigma$ generates positive mass
modulations with the desired characteristics. 
This realization demonstrates that the pulse hierarchy can arise from an
explicit local multifield dynamics.

\smallskip

We then apply the filter to a minimal spectator-dark-matter scenario, in
which isocurvature fluctuations get suppressed at large scales.  We
assume that the homogeneous spectator displacement is already present during
inflation, that the Coulomb stage terminates with inflation, and that the
spectator mass subsequently approaches a constant value $m_\sigma$.  The
field later begins coherent oscillations when the Hubble rate becomes of order
$m_\sigma$ and contributes dark matter through the usual quadratic
misalignment mechanism.  In this completion the turnover scale of the inflationary
spectator spectrum, represented in the lower right panel of Fig.~\ref{fig:single-positive-kick-summaryV1}, is related with $m_\sigma$, 
and provide the main prediction of our dark-matter
realization. A measurement or constraint on the isocurvature turnover becomes
a direct statement about the spectator mass relative to the inflationary
scale.
The  minimal realization is sufficiently restrictive that its consistency
conditions can be analyzed explicitly, by making use of our analytical
formula for the spectator spectrum. We exactly compute
conditions to ensure that backreaction of the spectator
dynamics can be set under control. At the observational level, we
impose a conservative CMB isocurvature requirement and compare the exact
smooth spectrum with existing constraints on blue and broken-power-law
cold-dark-matter isocurvature from the CMB, Lyman-$\alpha$ forest, spectral
distortions, and high-redshift galaxy luminosity functions --
we rely on the recent
\cite{Buckley:2025zgh,Co:2026klr}. The results favour scenarios with relatively
low inflationary scale.

The paper is organized as follows.
Section~\ref{sec:pulse-motivation} discusses physical mechanisms that can
produce short inflationary events and explains how they induce spectator-mass
kicks.  Section~\ref{sec:coulomb-filter-theory} develops the finite-pulse
transfer-matrix description, constructs the logarithmic tower, derives its
Coulomb continuum limit and exact transfer function, and discusses EFT control
and the massive-clock realization.  Section~\ref{sec:filtered-spectator-dm}
embeds the filter in spectator dark matter and studies the abundance,
homogeneous backreaction, fluctuation variance, observational constraints, and
representative benchmarks.  We conclude by placing the mechanism in the
broader context of blue-isocurvature models and by discussing its limitations
and possible extensions in Section \ref{sec:outlook}.  The appendices collect the finite-pulse transfer
matrices and the derivation of the universal Coulomb connection coefficient
needed in the main text.

\section{Physical motivations for inflationary pulses}
\label{sec:pulse-motivation}

Inflation is often treated as a slowly evolving background, lasting
at least sixty e-folds of expansion. But brief, controlled
departures from an adiabatic attractor evolution arise in many well-motivated models.
Sharp features in a potential, transitions between dynamical regimes,
turns in a multifield trajectory, or short episodes of particle production
can all generate localized time dependence in the equations governing
cosmological perturbations.  When the duration of such an event is short
compared with the characteristic time scale of the modes of interest, its
leading effect can be represented by  instantaneous pulses -- or kicks
 -- which as we shall learn pave
the way to an exactly solvable system with compelling   properties.

\smallskip

The purpose of this section is to motivate, and briefly explain, possible  physical origins for short-duration pulses.  What makes such setup particularly interesting is the possibility  
to change the amplitude of cosmological fluctuations at the scales where the pulse
occurs.
The most familiar examples in the literature
concern the dynamics of curvature perturbation, and we briefly review
them in Section \ref{subsec:curvature-pulses}. But for our purposes, we plan to develop the idea 
in a spectator sector, while the adiabatic inflationary evolution remains
standard (see Section \ref{subsec:spectator-mass-kick-origin}).

\subsection{Temporal scattering in the curvature sector}
\label{subsec:curvature-pulses}

For a broad class of single-clock scalar theories, the quadratic action for
the curvature perturbation is
\begin{equation}
 S_\zeta
 =
 \frac12\int d\tau\,d^3x\,
 z^2(\tau)
 \left[
   \zeta'^2-c_s^2(\tau)(\boldsymbol\nabla\zeta)^2
 \right],
 \label{eq:pulse-zeta-action}
\end{equation}
where, in the simplest effective-field-theory normalization,
\begin{equation}
 z^2=\frac{2a^2\epsilon M_{\rm Pl}^2}{c_s^2},
 \qquad
 \epsilon\equiv-\frac{\dot H}{H^2}.
 \label{eq:pulse-z-definition}
\end{equation}
and $\zeta$ is the curvature perturbation \cite{Liddle:2000cg}. 
The canonical variable \(v=z\zeta\) satisfies
\begin{equation}
 v_k''
 +
 \left[
   c_s^2 k^2-\frac{z''}{z}
 \right]v_k=0.
 \label{eq:pulse-MS}
\end{equation}
A rapid change in the background, encoded in the so-called
pump function \(z(\tau)\),  acts as a time-dependent
scattering potential for \(v_k\).

\smallskip
Several established mechanisms exemplify how localized time dependence can
arise:
\begin{itemize}
\item{\bf Sharp features in the inflaton potential.}
Consider an inflationary potential of the form
\begin{equation}
 V(\phi)
 =
 V_{\rm sr}(\phi)
 +
 \Delta V\,
 {\cal S}\!\left(\frac{\phi-\phi_\star}{d}\right),
 \label{eq:pulse-step-potential}
\end{equation}
where the function \({\cal S}\) changes appreciably over a short field interval \(d\), while
$\Delta V$ is a dimensionful quantity controlling the pulse amplitude.
As the inflaton crosses the feature at around $\phi_\star$, its acceleration 
varies over a short time scale
\begin{equation}
 \Delta t\sim\frac{d}{|\dot\phi_\star|}.
 \label{eq:pulse-step-duration}
\end{equation}
Even when \(\Delta V/V\ll1\), a sufficiently small width \(d\) can produce
a short violation of slow roll and a localized structure in \(z''/z\) 
\cite{Starobinsky:1992ts,Leach:2001zf}. 
Sharp steps generate oscillatory features in the scalar power spectrum and
correlated signatures in higher-point functions
\cite{Adams:2001vc,Adshead:2011jq}.  They provide the simplest example of a
small disturbance in the background energy density whose  time derivatives become
nevertheless large for a brief amount of time time, and
have been explored in classic model building works

\item{\bf Transient non-attractor evolution.}
A sufficiently flat portion of the inflationary potential can produce an
ultra-slow-roll phase,
\begin{equation}
 \ddot\phi+3H\dot\phi\simeq0,
 \qquad
 \dot\phi\propto a^{-3},
 \qquad
 \epsilon\propto a^{-6}.
 \label{eq:pulse-USR}
\end{equation}
The transitions into and out of this phase generate localized changes in
the curvature pump function $z(\tau)$.  In addition, the second long-wavelength solution for
\(\zeta\) is no longer negligible, and the curvature perturbation can evolve
outside the horizon.  A finite
non-attractor episode naturally contains two edges:
 {attractor}, 
 {non-attractor}, 
 {attractor again},
which become two separated pulses in a sufficiently sharp description.
Such constructions are well developed in the physics of primordial
black hole model building, see e.g. \cite{Ozsoy:2023ryl} for a theoretically
oriented review. 

\item{\bf Turns in a multifield trajectory.}
Suppose that the inflationary background follows a curved trajectory in
field space with angular velocity \(\dot\theta\).  Suppose also that  the orthogonal
fluctuation is sufficiently heavy,  and remains adiabatically decoupled.
Integrating it out gives, schematically, the estimate 
\begin{equation}
 c_s^{-2}-1
 \simeq
 \frac{\dot\theta^2}{M_{\rm eff}^2}.
 \label{eq:pulse-turn-cs}
\end{equation}
A localized turn can therefore induce simultaneous features in \(c_s\),
\(z\), and \(z''/z\), together with oscillatory imprints in the correlation
functions \cite{Achucarro:2010da,Achucarro:2012sm,Ozsoy:2018flq}.  The reduced
single-field description requires generalized adiabaticity conditions,
which may be summarized as ($\omega$ being the characteristic mode frequency)
\begin{equation}
 \left|
 \frac{\dot\omega_{\rm heavy}}
      {\omega_{\rm heavy}^2}
 \right|
 \ll1.
 \label{eq:pulse-heavy-adiabaticity}
\end{equation}
which ensures that our split of background versus fluctuations is consistent. 
If this condition fails, the heavy mode must be retained explicitly: the
short event would then correspond to a genuine multifield excitation or particle
production, rather than merely to a transient sound speed.

\item{\bf Waterfall transitions and symmetry breaking.}
In hybrid-type inflationary models, a field \(\chi\) can undergo a rapid change of
stability when its effective mass crosses zero.  Couplings such as
\begin{equation}
 g^2\phi^2\chi^2
\end{equation}
then modify the scalar mass matrix over a short interval.  Waterfall
dynamics provide explicit examples in which a phase transition during
inflation generates strongly scale-dependent perturbations
\cite{Abolhasani:2010kr,Lyth:2012yp}.  A sharp effective description may
then be
useful at the linearized level --  although genuinely tachyonic transitions can
also excite nonlinear fluctuations that are not captured by simple
delta-function pulses.

\item{\bf Particle-production events and periodic structure.}
A rolling inflationary background might repeatedly cross enhanced-symmetry points or
regions in which an additional field becomes nonadiabatic.  A prototype
interaction is
\begin{equation}
 {\cal L}_{\rm int}
 =
 -\frac{g^2}{2}
 (\phi-\phi_j)^2\chi^2.
 \label{eq:pulse-particle-production}
\end{equation}
Near the positions \(\phi=\phi_j\), the \(\chi\) mass changes rapidly and a burst of
particle production can occur.  One or several such events can generate
localized features \cite{Kofman:2004yc,Barnaby:2009dd,Green:2009ds,Pearce:2017bdc}.  Periodic structures in field
space, including modulated axion potentials, provide a natural route to
repeated events during inflation
 \cite{Flauger:2009ab,Flauger:2010ja,Behbahani:2011it}.  If the background
advances approximately uniformly per e-fold, a periodic sequence in field
space becomes approximately equally spaced in the number of e-folds. We should keep this idea in mind, since we are going to use it in building an explicit realization of our filter 
mechanism with spectator fields.

\end{itemize}

To summarize,
a  pulse contribution to the pump-field combination  \(z''/z\)  can originate
from the dynamics of \(a\), \(\epsilon\), or \(c_s\). The  same transient
evolution generally modifies cubic and higher-order interactions.  We can
consequently regard
the
delta-function as a useful analytic idealization --  but a
complete curvature-sector model must also carefully control the background evolution,
non-Gaussianity, and the validity of the effective theory.  Avoiding such complications 
is one reason why in the present work we are going to apply  the pulse construction to a
spectator field, rather than directly to the adiabatic curvature fluctuation.

\subsection{From a rapid background event to a spectator-mass kick}
\label{subsec:spectator-mass-kick-origin}

The preceding examples indicate that short, localized changes of
inflationary parameters are in fact common in model-building literature.  In view of our dark matter applications, we
use a specific implementation: we assume the inflaton background and its
adiabatic perturbation remain  standard, while some  mechanisms~\footnote{We
present an explicit example in Section~\ref{subsec:massive-clock-realization}.}
 influence the mass of a spectator scalar field \(\sigma\).

\smallskip

A generic quadratic-type interaction of the type we are interested in may be written as
\begin{equation}
 V_{\rm int}(\sigma,\tau)
 =
 \frac12\,  M_\sigma^2(\tau)\,\sigma^2,
 \label{eq:pulse-spectator-portal}
\end{equation}
where \(\tau\) denotes conformal time. Such a time-dependent mass might
be induced by couplings with 
the inflaton itself or an extra independent
subdominant field.  A brief  sizeable excursion of \( M_\sigma\) around a time \(\tau_j\)
produces a time-dependent mass contribution  \( M_{\sigma,j}^2\)  localized over a short conformal-time interval
\(\Delta\tau_j\).

Focussing on  the canonical spectator variable \(u=a\sigma\), we  consider a mode
 equation expressed  as
\begin{equation}
 u_k''
 +
 \left[
   k^2-\frac{a''}{a}+a^2\, M_\sigma^2(\tau)
 \right]u_k=0.
 \label{eq:pulse-spectator-mode}
\end{equation}
with a time-dependent mass only turned on around the pulses at  \(\tau_j\). 
When the spectator is massless, and no pulses are present, the positive-frequency solution
of Eq.~\eqref{eq:pulse-spectator-mode}  matching
the Bunch-Davies condition (for $a(\tau) =-1/(H_{\rm inf} \tau)$)  is
\begin{equation}
 f_k(\tau)=\frac{1}{\sqrt{2k}}
 \left(1-\frac{i}{k\tau}\right)e^{-ik\tau},
 \label{eq:massless-bd-mode}
\end{equation}
which gives the standard expression
for the power spectrum of the spectator $\sigma$ \cite{Liddle:2000cg}
\begin{equation}
 \mathcal P_{\sigma}^{(0)}(k)
 =\left(\frac{H_{\rm inf}}{2\pi}\right)^2.
 \label{eq:massless-spectator-power}
\end{equation}
But, more generally, the mass term in Eq.~\eqref{eq:pulse-spectator-mode}
changes this simple result.  
If the instantaneous  event around $\tau_j$ is unresolved by the modes of interest -- we discuss  in Section~\ref{sec_radcor} the corresponding conditions --  its leading effect
depends only on its integrated area,
\begin{equation}
 s_j
 \equiv
 \int_{\tau_j-\Delta\tau_j/2}^{\tau_j+\Delta\tau_j/2}
 d\tau\,a^2(\tau)\, M_{\sigma,j}^2(\tau).
 \label{eq:pulse-area}
\end{equation}
One may therefore replace the finite profile by an effective $\delta$-function description 
\begin{equation}
 a^2 M_{\sigma,j}^2(\tau)
 \ \longrightarrow\
 s_j\,\delta(\tau-\tau_j).
 \label{eq:pulse-delta-limit}
\end{equation}
Integrating Eq.~\eqref{eq:pulse-spectator-mode} across these events gives
Israel-type conditions \cite{Israel:1966rt}
\begin{equation}
 [u_k]_j=0,
 \qquad
 [u_k']_j=-s_j u_k(\tau_j)\,,
 \label{eq:pulse-jump-condition}
\end{equation}
with $[\dots]_j$ indicating the difference of the evaluated quantities among
the two temporal sides of the event.
A positive excursion of the spectator mass corresponds to
an integrated area \(s_j>0\).

Once we take into 
 due account the contributions of the temporal pulses,  we evaluate
 the spectrum towards the end of inflation, and
 formally write it  as
\begin{equation}
 \mathcal P_{\sigma}(k)
 =\left(\frac{H_{\rm inf}}{2\pi}\right)^2\mathcal T(k),
 \label{eq:transfer-definition}
\end{equation}
where $\mathcal T=1$ in the absence of a time-dependent mass. 
The function $\mathcal T(k)$ is a transfer function encapsulating
the effects
of temporal kicks which modulate the spectrum between large towards small scales.
Transient pulses, as we shall see, are able to considerably change the amplitude of $ \mathcal P_{\sigma}(k)$ between different scales.

\smallskip

The spectator implementation in terms of the field $\sigma$
has the advantage of keeping the system compatible with an approximately standard inflaton adiabatic sector,    provided that
the energy stored in the spectator and in any produced
quanta remains subdominant.  These requirements, if satisfied, have interesting
phenomenological ramifications, which we are going to explore.

\section{Inflationary mass kicks and the Coulomb filter}
\label{sec:coulomb-filter-theory}

We  examine a  spectator $\sigma$ whose background energy density is
subdominant during inflation.  Inflation itself is therefore kept close to de Sitter, with
scale factor $a(\tau) =-1/(H_{\rm inf} \tau)$ and $\tau<0$. 
 We
show how a system of well-located
$\delta$-function contributions to the spectator mass -- which we call
 kicks, or pulses --  can modulate
and change the amplitude of the spectrum 
of fluctuations of the spectator field,  
leading to a structure with interesting analytic properties, special 
features for what respect its radiative corrections,  and convenient
applications for the physics of the dark sector.

\smallskip
In this Section
 we
 develop a sequence of increasingly smooth
filters: 

\begin{itemize}
\item[i)] one  kick pulse, and two-kick interference: Section \ref{subsec:isolated-kicks-summary}. These
 systems provide intuition on how the pulses
 change the spectator spectrum profile, the physics of temporal scattering, 
  and allows us to fix the matching
conventions;
 \item[ii)]  a logarithmic tower, an example of  system 
where pulse positions accumulate towards a fixed point. Section \ref{subsec:logarithmic-tower} 
 develops such a framework, and motivates it in terms of a discrete self-similar
 condition;
  \item[iii)]  the  Coulomb continuum limit of the logarithmic
tower, leading to a smooth analytic spectrum of fluctuations with   a well-defined turnover scale: Section \ref{subsec:exact-coulomb-filter}. This profile
of the spectrum has interesting phenomenological
consequences, which are important for our applications to dark matter.
\end{itemize}
After establishing the general properties of the setup, we also examine
some of its theoretical ramifications:
\begin{itemize}
\item[iv)]  Section \ref{sec_radcor}  discusses features of radiative corrections in this scenario, which depends on the symmetry properties of the system;
 \item[v)]  in Section \ref{subsec:massive-clock-realization} we  examine  a possible realization of the idea in terms of a massive
 clock during inflation.
\end{itemize}

\subsection{What one and two kicks during inflation teach us}
\label{subsec:isolated-kicks-summary}

We first
consider  a finite set of instantaneous positive mass pulses, with the choice
\begin{equation}
 a^2\, M_\sigma^2(\tau)=\sum_j s_j\,\delta(\tau+T_j),
 \qquad s_j>0,
 \qquad T_j>0.
 \label{eq:finite-kick-profile}
\end{equation}
in Eq. ~\eqref{eq:pulse-spectator-mode}.
From now on, we indicate with a capital $T$ positive quantities as $T=-\tau$ during inflation (since $\tau<0$). The quantities $s_j$ above correspond to each pulse-integrated
area -- recall the discussion around Eq.~\eqref{eq:pulse-delta-limit}.

At the $j$th pulse, located at $\tau_j=-T_j$, the canonical mode obeys
Israel boundary conditions
\begin{equation}
 [u_k]_j=0,
 \qquad
 [u_k']_j=-s_j u_k(\tau_j).
 \label{eq:kick-jump-conditions}
\end{equation}
Starting from a mode function in Bunch-Davies vacuum $f_k(\tau)$, 
see Eq.~\eqref{eq:massless-bd-mode}, 
the first pulse acts as temporal scatterer: it switches on  reflection and transmission effects controlled
by Bogoliubov coefficients $(\alpha,\,\beta)$ leading to a mode function $u_k^{(0)}= \alpha_1\, f_k+\beta_1 \, f_k^{*}$ \cite{Birrell:1982ix}.
The exact transfer matrices and the corresponding one- and two-kick
Bogoliubov coefficients are given in
Appendix~\ref{app:one-two-kicks}.  Here we retain only the physical lessons
that motivate the continuum construction we develop next.

\smallskip

For a single kick at conformal time $\tau=-T$, we define
the dimensionless combinations
\begin{equation}
 x\equiv kT,
 \qquad
 q\equiv sT.
 \label{eq:single-kick-variables}
\end{equation}
After the pulse, 
 the late-time transfer
function of Eq.~\eqref{eq:transfer-definition} reads
\begin{equation}
 \mathcal T_1(k)=|\alpha_1-\beta_1|^2.
 \label{eq:single-kick-transfer}
\end{equation}
At the phenomenological level, the most interesting regime is the so-called
 near-critical positive kick
\begin{equation}
 q=3-\delta,
 \qquad 0<\delta\ll1,
 \label{eq:single-kick-near-critical}
\end{equation}
which requires us to fine-tune the parameter $q$ introduced in Eq.~\eqref{eq:single-kick-variables}. This condition 
identifies a regime  leading to a  hierarchy
in amplitude between the spectator sector at large ($x\ll1$) versus small
($x\gg 1$) scales. Hence the system acts as a filter
suppressing the large-scale spectrum in comparison with the small-scale
one.

\begin{figure}[t]
 \centering
   \includegraphics[width=0.6\linewidth]{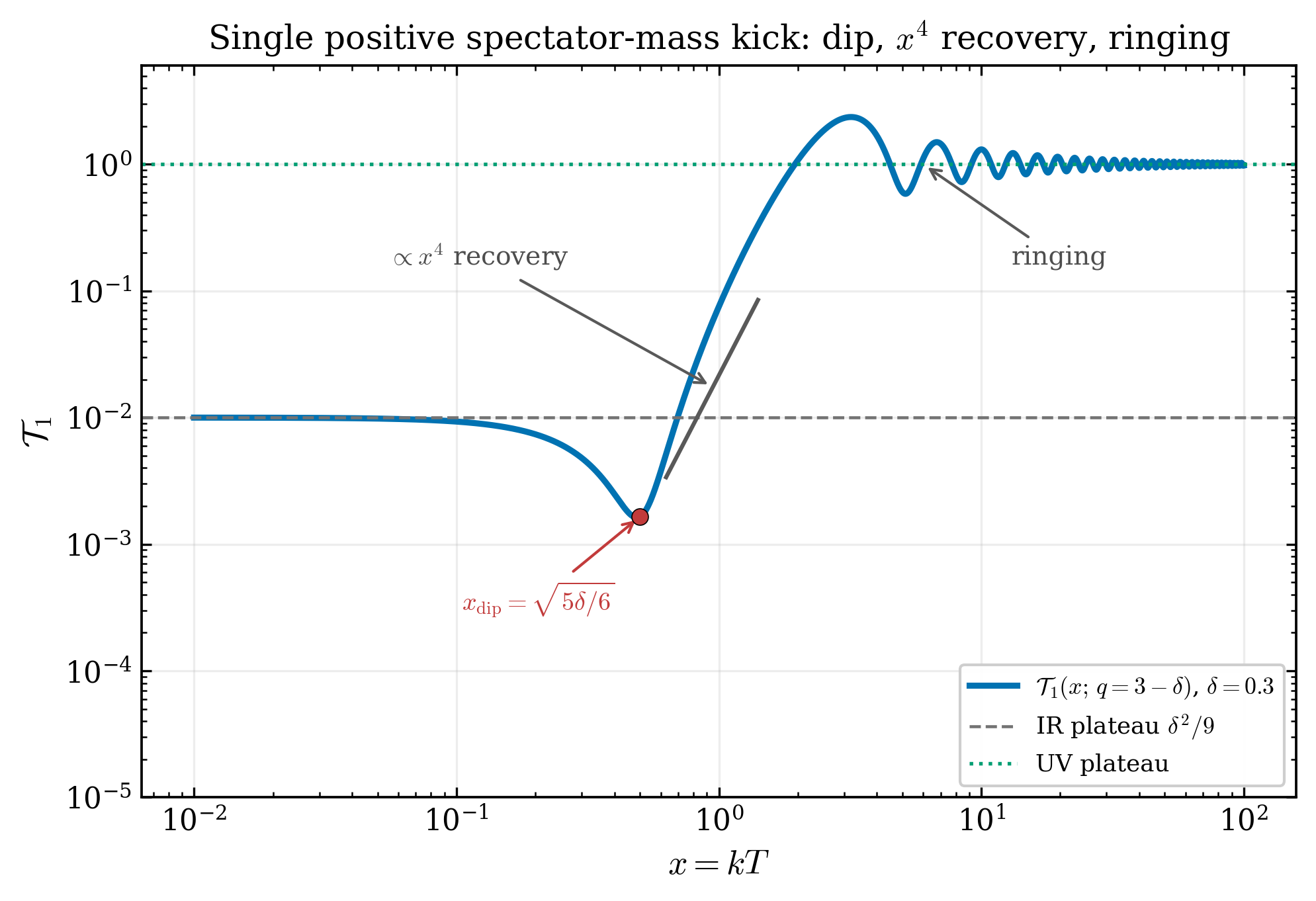}
 \caption{\small
  Transfer function $\Tcal_1(x;q)$ of a single positive spectator-mass kick,
  for $q=3-\delta$ with $\delta=0.3$, as a function of $x=kT$ with $T=-\tau$
  the conformal time of the kick.  Modes far outside the horizon at the time
  of the kick are suppressed to the infrared plateau $\delta^2/9$, while modes
  already deep inside the horizon are unaffected and approach the ultraviolet
  plateau $\Tcal_1\to1$.  Between the two the response passes through a zero of
  the leading long-wavelength amplitude at
  $x_{\rm dip}=\sqrt{5\delta/6}$ (marked), recovers as $x^4$, and then rings
  with period $\Delta x=\pi$ about the ultraviolet plateau, the ringing being
  the imprint of the sharp switch-on.}
 \label{fig:single-positive-kick-summary}
\end{figure}

In fact, this limit leads to   a useful elementary high-pass filter. At small $x$ we find
\begin{equation}
 \mathcal T_1(x)
 \simeq
 \left(\frac{\delta}{3}-\frac{2x^2}{5}\right)^2,
 \label{eq:single-kick-small-x}
\end{equation}
so that
\begin{equation}
 \mathcal T_1(0)=\frac{\delta^2}{9},
 \qquad
 x_{\rm dip}^2\simeq\frac{5\delta}{6},
 \qquad
 \mathcal T_1\simeq\frac{4x^4}{25}
 \quad(\delta\ll x^2\ll1),
 \label{eq:single-kick-regimes-compact}
\end{equation}
while $\mathcal T_1\to1$ for $x\gg1$.  A single positive kick localized in time therefore
suppresses sufficiently long modes: it leaves a localized dip, followed
by a $x^4$ growth,  and coherent
ringing around $k\sim T^{-1}$ leading to a flat envelope. See Fig.~\ref{fig:single-positive-kick-summary}. The growth
with slope $x^4$ visible in the plot is in fact a {\it universal feature} of a system
with short violations of slow-roll evolution, and very well studied in the context
of primordial black hole model building \cite{Byrnes:2018txb,Carrilho:2019oqg,Ozsoy:2019lyy,Cole:2022xqc}  \footnote{In fact, also in  primordial black hole physics we are interested in models
with a spectrum of fluctuations rapidly raising from large towards small scales. }, a setup with borrows some analogy with the present framework. 
 We regard this single-pulse system
as an instructive prototype, but not yet leading to 
broad, simple and smooth filter, because of  the dip and the ringing features
visible in
 Fig.~\ref{fig:single-positive-kick-summary}, which introduce several distinct
 scale in the problem (but see \cite{Tasinato:2020vdk,Tasinato:2023ukp} for a method to relate them). One of our aims 
 is to determine a simpler, exact spectrum characterized by a single turnover scale only,
 which will allow us to gain  analytical control of the properties of the system when applied
 to dark matter model building. 

\smallskip

With two kicks, the second pulse acts on the squeezed state created by the
first, and leads to  interference phenomena, which
can be carefully exploited
to modulate the spectrum.  The result of applying two pulses is expressed as an ordered product of transfer matrices $\mathsf K_i$ -- see the discussion
 in Appendix \ref{app:one-two-kicks}
\begin{equation}
 \begin{pmatrix}\alpha_{12}\\ \beta_{12}\end{pmatrix}
 =\mathsf K_2\mathsf K_1
 \begin{pmatrix}1\\0\end{pmatrix},
 \qquad
 \mathcal T_2(k)=|\alpha_{12}-\beta_{12}|^2.
 \label{eq:two-kick-product-summary}
\end{equation}
For weak  pulses, we find that the second Bogoliubov coefficient $\beta$, typically
associated with particle creation, results 
\begin{equation}
 |\beta_{12}|^2\simeq
 \frac{s_1^2+s_2^2+2s_1s_2
 \cos[2k(T_1-T_2)]}{4k^2}.
 \label{eq:two-kick-interference-summary}
\end{equation}
The pulse separation controls the oscillation period, while the relative
areas control the interference contrast.  This phase-coherent multiplication
is an essential feature that persists for a large number of kicks.
When the position and amplitude of the pulses is appropriately
tuned, interference effects can be combined,   giving interesting
physical ramifications. 

\subsection{A logarithmic tower of pulses, and its continuum limit}
\label{subsec:logarithmic-tower}

Is it possible to distribute the pulses in a symmetrical, convenient
way so to improve  the properties of the spectator spectrum? We
do so in this section, distributing $N$ positive pulses over a finite interval of conformal
time, and taking $N$ large. Our aim is to find a system
that modulates appropriately the spectrum
of spectator-field fluctuations leading a simple analytical configuration
with a universal power-law growth in the infrared and a single turnover scale.  (Recall  Fig.~\ref{fig:single-positive-kick-summaryV1}.)

 Let
\begin{equation}
 T_j=T_{\rm on}e^{-jh},
 \qquad
 j=0,\ldots,N-1,
 \qquad
 T_{\rm off}\simeq T_{\rm on}e^{-Nh},
 \label{eq:log-pulse-times}
\end{equation}
where $h>0$ controls  the pulse separation in $\ln T$. In conformal time, the pulses
accumulate towards $|\tau|\to0$.  Since $a\propto T^{-1}$ in de
Sitter, adjacent pulses are equally spaced in e-folds,
\begin{equation}
 \Delta N=h.
 \label{eq:pulse-spacing-efolds}
\end{equation}
making the condition \eqref{eq:log-pulse-times} physically more transparent. In fact, 
a succession of equally-spaced kicks can be realized
in systems with periodically modulated
potentials -- see  the explicit realization to be developed in Section \ref{subsec:massive-clock-realization}. 
We
choose a constant pulse area 
\begin{equation}
 s_j=h\kappa_c,
 \label{eq:constant-pulse-area}
\end{equation}
for each $j$
in Eq \eqref{eq:finite-kick-profile},
where $\kappa_c$ has dimensions of comoving momentum. The
resulting setup is  characterized by a self-similar configuration
with discrete translational invariance, 
 since the system  is invariant under rescaling $\tau
\to e^h\,\tau$, combined with translations $j\to j+1$. Moreover, it is also
equipped by a shift symmetry $\sigma \to \sigma+c$, broken only 
at the positions $\tau_i$ of the pulses.
 These properties
play a role  for characterizing the radiative stability of the framework,
see Section \ref{sec_radcor}.  As in the (fine-tuned) single
kick event of Section \ref{subsec:isolated-kicks-summary}, the concerted action of many pulses located 
appropriated lead to a suppression of large-scale spectrum with respect
to the small-scale one, as we are going to discuss. 

\smallskip

  In the dense limit, 
the sum over pulses becomes a Riemann integral, and we can coarse-grain the system.  Since, formally, 
\begin{equation}
 dj\,=\,-\frac{dT}{h\,T},
 \label{eq:pulse-density}
\end{equation}
we
can take a continuum limit of the dense spectrum (recall $ T=-\tau$)
\begin{equation}
J(\tau)\equiv
 \sum_j h\kappa_c\,\delta(T-T_j)
 \;\longrightarrow\;
 \frac{\kappa_c}{T}
 \Theta(T_{\rm on}-T)\Theta(T-T_{\rm off})\,.
 \label{eq:finite-window-continuum}
\end{equation}
Thus the log-spaced tower, denoted with the combination $J(\tau)$ above,  coarse-grains to a simple  expression
\begin{equation}
 a^2\,\delta M_\sigma^2(\tau)
 \,=\,\frac{\kappa_c}{-\tau}
 \Theta(T_{\rm on}+\tau)\Theta(-\tau-T_{\rm off}).
 \label{eq:finite-coulomb-window}
\end{equation}
The ideal analytic filter is obtained by extending the window over the range
relevant to the inflationary modes under consideration. In this window, $T_{\rm on}\le -\tau \le T_{\rm off}$.  Hence, within a large window, we obtain
a single-parameter expression (see Fig.~\ref{fig:log-tower-continuum}, left panel) 
\begin{equation}
 {
 a^2M_\sigma^2(\tau)=\frac{\kappa_c}{-\tau}.}
 \label{eq:ideal-coulomb-profile}
\end{equation}

\paragraph{Discrete vs Continous} The consequences of
equation~\eqref{eq:ideal-coulomb-profile} may therefore be viewed in two
complementary ways, a discrete and a continuous one.   It is the dense limit of repeated kicks, and it is also
a continuous mass profile in its own right.  The first interpretation makes
the scattering origin of the system transparent and it points towards a definite microscopic description with interesting properties under radiative corrections, 
and for possible embedding in massive-clock inflationary scenarios (see 
Sections \ref{sec_radcor} and \ref{subsec:massive-clock-realization}).
The second, coarse-grained  perspective gives a simple analytic effective
model with a specific time-dependent mass -- which we are now going to analyze and solve exactly. In what follows, we 
will often compare the two viewpoints and adopt the most convenient one depending
on circumstances. 

\subsection{The exactly soluble Coulomb filter}
\label{subsec:exact-coulomb-filter}

For the smooth profile in Eq.~\eqref{eq:ideal-coulomb-profile}, the mode
equation  becomes
\begin{equation}
 u_k''+
 \left[k^2-\frac{2}{\tau^2}+\frac{\kappa_c}{-\tau}\right]u_k=0.
 \label{eq:coulomb-mode-tau}
\end{equation}
Using the previous dimensionless
quantity $x\equiv-k\tau>0$,  we obtain
\begin{equation}
 \frac{d^2u_k}{dx^2}
 +\left[1+\frac{\kappa_c}{k\,x}-\frac{2}{x^2}\right]u_k=0.
 \label{eq:coulomb-mode-x}
\end{equation}
This is the standard radial Coulomb equation
\begin{equation}
 u_{xx}+\left[1-\frac{2\eta}{x}
 -\frac{\ell(\ell+1)}{x^2}\right]u=0\,
 \label{eq:standard-coulomb-equation}
\end{equation}
when selecting $
 \ell=1$, 
$ \eta(k)=-{\kappa_c}/{(2k)}$.
 The problem at
hand
becomes conceptually similar to a Coulomb scattering system well-studied in quantum mechanics,
and we can build on well-established, classic 
results~\footnote{The Coulomb equation has already been explored in
inflationary literature in other contexts, see e.g. \cite{Martin:2000ei}.}.

\begin{figure}[t!]
 \centering
 \IfFileExists{fig_pulse_tower_to_coulomb_profile.png}{%
 \includegraphics[width=0.49\linewidth]
 {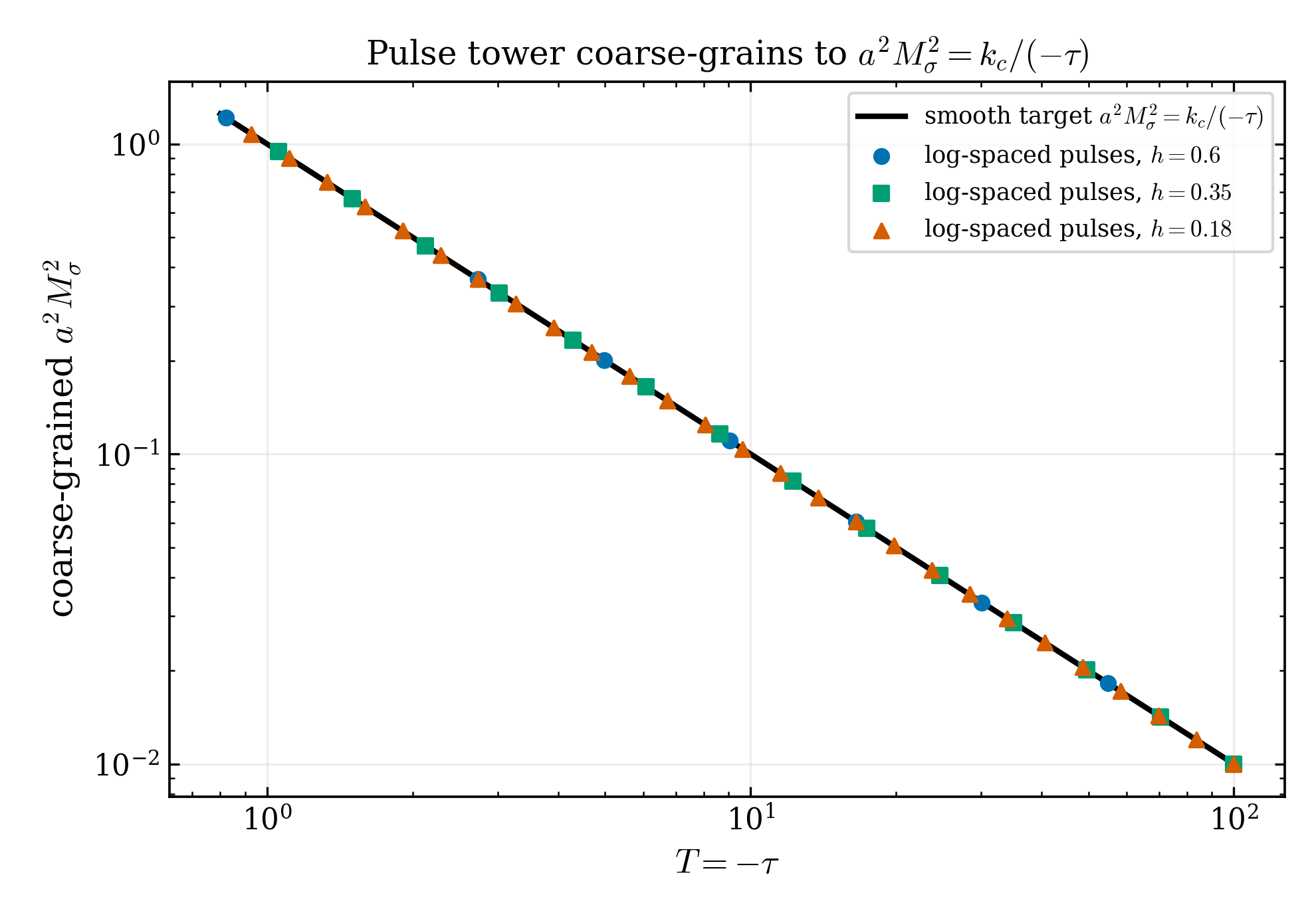}}{%
 \fbox{\parbox[c][4.5cm][c]{0.72\linewidth}{\centering
 Placeholder for the logarithmic pulse tower and its continuum profile.}}}
 \IfFileExists{fig_coulomb_filter_transfer_universal.png}{%
 \includegraphics[width=0.49\linewidth]
 {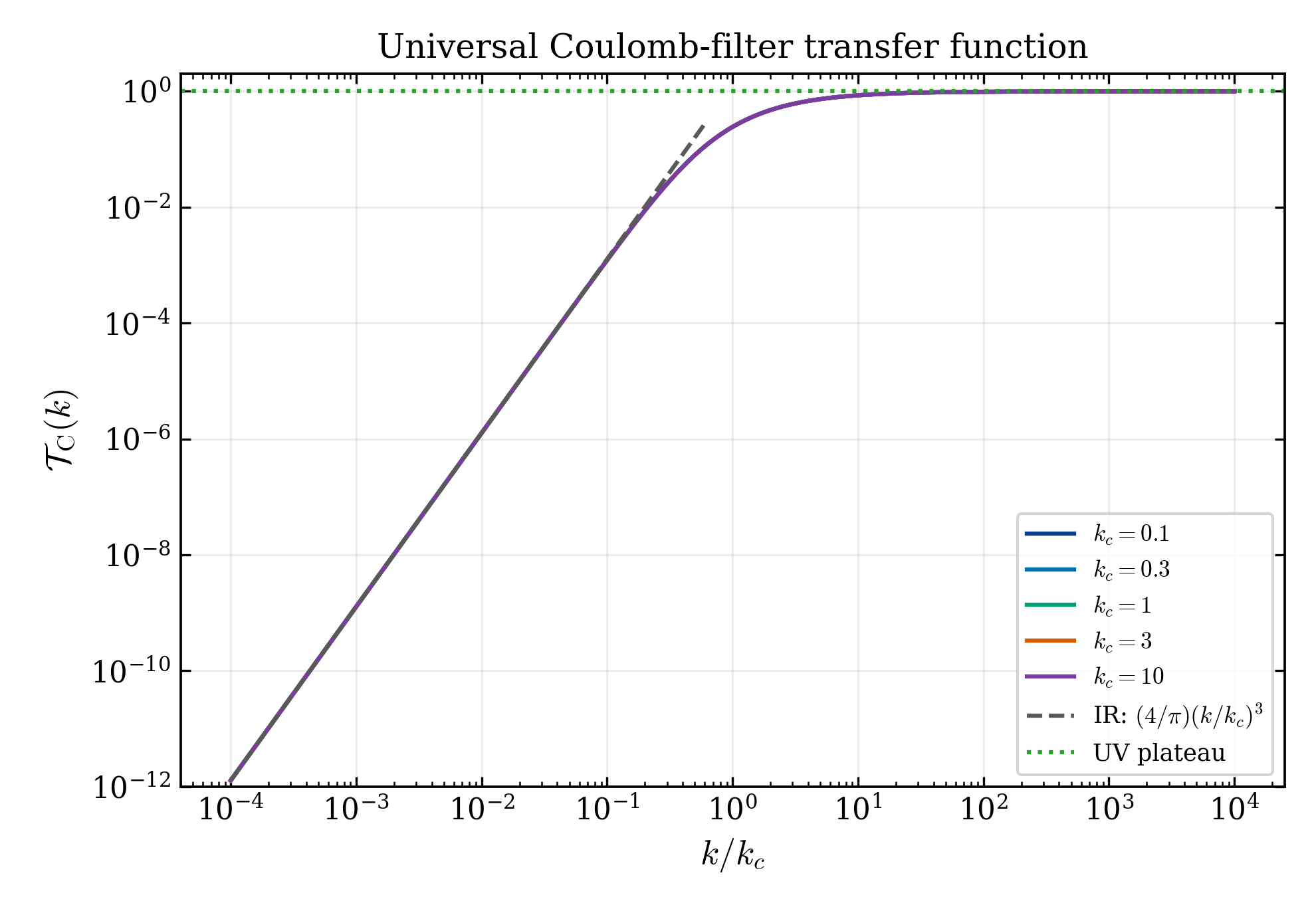}}{%
 \fbox{\parbox[c][4.5cm][c]{0.72\linewidth}{\centering
 Placeholder for the universal Coulomb-filter transfer function.}}}
 \caption{\small {\bf Left panel:} Coarse graining of a logarithmically spaced pulse tower.  A pulse
 area $s_j=h\kappa_c$ divided by the local bin width
 $\Delta T_j\simeq hT_j$ gives the smooth envelope
 $a^2M_\sigma^2\simeq\kappa_c/T$. {\bf Right panel:} Universal Coulomb transfer function,
 Eq.~\eqref{eq:coulomb-transfer}.  Long modes are suppressed as $k^3$, while
 short modes approach the unfiltered spectator spectrum.}
 \label{fig:log-tower-continuum}
\end{figure}

\smallskip

The
solution can be expressed through Coulomb functions $F_1(\eta,x)$ and $G_1(\eta,x)$, or through
Whittaker functions.  We provide
all details of the calculations in Appendix \ref{sec:coulomb-transfer-derivation}. By selecting the positive-frequency Coulomb wave at early
times -- so to match with the standard Bunch-Davies condition -- and by extracting the constant late-time mode, we obtain  the universal
transfer function 
\begin{equation}
 {
 \mathcal T_{\rm C}(k)=
 \frac{1-e^{-\pi\kappa_c/k}}
 {(\pi\kappa_c/k)\left[1+\kappa_c^2/(4k^2)\right]}.}
 \label{eq:coulomb-transfer}
\end{equation}
which enters the spectator spectrum $\mathcal P_{\sigma}(k)=
 \left({H_{\rm inf}}/{2\pi}\right)^2
 \mathcal T_{\rm C}(k)$ as in Eq.~\eqref{eq:transfer-definition}.
The transfer function has a smooth, regular profile without any ringing behaviour,
nor dip features, and depends
on a single scale $\kc$. See Fig.\ref{fig:log-tower-continuum}, right panel. The simple analytical expression \eqref{eq:coulomb-transfer} is one of the main results of this work. Interference
effects  of many log-spaced pulses compensate each other giving the previous simple expression for small pulse separation $h$, which
depend on a single  scale $\kappa_c$ without further tunings.  Two limits  make the filtering action explicit:
\begin{align}
 \mathcal T_{\rm C}(k)
 &\simeq
 \frac{4}{\pi}\left(\frac{k}{\kappa_c}\right)^3,
 && k\ll\kappa_c,
 \label{eq:coulomb-transfer-ir}
 \\
 \mathcal T_{\rm C}(k)
 &=1-\frac{\pi\kappa_c}{2k}
 +\mathcal O\!\left(\frac{\kappa_c^2}{k^2}\right),
 && k\gg\kappa_c.
 \label{eq:coulomb-transfer-uv}
\end{align}
Thus the  free parameter $\kappa_c$ sets the turnover scale.
  Modes
with $k\ll\kappa_c$ acquire a strongly blue spectrum,
$\mathcal P_{\sigma}\propto k^3$, whereas modes with
$k\gg\kappa_c$ retain the usual massless-spectator amplitude. 
The turnover
quantity $\kappa_c$ has  a transparent physical interpretation: 
setting $M_\sigma = H$ in the expression $a^2 M_\sigma^2 = \kappa_c/(-\tau)$, the scale $\kappa_c$ corresponds to the comoving
scale leaving the horizon at mass-Hubble crossing. Modes
that leave the horizon sufficiently early have large mass with respect to Hubble, hence their
fluctuations are suppressed. Modes that cross the horizon later
during inflation have small mass, hence they keep scale invariant.
Recall Fig.~\ref{fig:single-positive-kick-summaryV1} in the Introduction, and its accompanying arguments.

\smallskip

 Ours is
 an explicit demonstration of a system attaining
the maximal infrared  slope allowed by causality, $\left({k}/{\kappa_c}\right)^3$,  in a relatively simple, analytical example.
  The smooth
nature of the spectrum will be helpful in our application to dark matter physics,
see Section \ref{sec:filtered-spectator-dm}.

\medskip

The exact expression~\eqref{eq:coulomb-transfer} describes the ideal profile,
acting over an unbounded range of conformal time.  A physical realization has a
finite onset $T_{\rm on}$ and a finite termination $T_{\rm off}$, introducing
two additional scales,
\begin{equation}
  k_{\rm on}\sim T_{\rm on}^{-1},
  \qquad
  k_{\rm off}\sim T_{\rm off}^{-1}.
  \label{eq:on-off-scales}
\end{equation}
A natural first guess is to demand that each mode of interest be both deep
inside the horizon and unaffected by the Coulomb term at the onset,
\begin{equation}
  x_{\rm on}=kT_{\rm on}\gg1,
  \qquad
  x_{\rm on}\gg\frac{\kappa_c}{k}.
  \label{eq:early-onset-condition}
\end{equation}
The second of these conditions deserves care, because it is not a single requirement but a
$k$-dependent one that degrades quadratically towards the infrared.  Writing
$x_{\rm on}=kT_{\rm on}$, it reads $k^{2}\gg\kappa_c/T_{\rm on}$, that is
$  k\;\gg\;\sqrt{{\kappa_c}/{T_{\rm on}}}$.
%
But
Eq.~\eqref{eq:early-onset-condition} is stronger than
necessary.  What the second condition really protects against is the
\emph{discontinuity} of $a^2M_\sigma^2$ at $T_{\rm on}$, not the presence of the
Coulomb term as such.  A mode that is deep inside the horizon remains
adiabatic at the onset even when the Coulomb term dominates over $k^2$: with
$\omega^2=k^2+\kappa_c/(-\tau)$ one finds
\begin{equation}
  \left|\frac{\omega'}{\omega^{2}}\right|
  \;\simeq\;
  \frac{1}{2}\,\frac{1}{\sqrt{x_{\rm on}\,\kappa_c/k}}
  \;\ll\;1 ,
  \label{eq:onset-adiabaticity}
\end{equation}
which is satisfied far more easily than
Eq.~\eqref{eq:early-onset-condition}.  The correct requirement is therefore not
that the Coulomb term be negligible at the onset, but that it be switched on
\emph{adiabatically}: writing $a^2M_\sigma^2=(\kappa_c/T)\,S(T)$ for a switching
envelope $S$ that rises from zero to unity over $w$ e-folds, we require
\begin{equation}
  w\;\gtrsim\;\mathcal{O}(1)\quad\text{e-folds},
  \label{eq:onset-adiabatic-condition}
\end{equation}
in place of the second condition in Eq.~\eqref{eq:early-onset-condition}.
%
In what follows we  assume
$x_{\rm on}\gg1$ together with an adiabatic onset, and use the ideal
expression~\eqref{eq:coulomb-transfer} throughout.  The termination of the
profile must likewise be matched to the post-inflationary evolution; we return
to this in Section~\ref{sec:outlook}.

\subsection{Behaviour under radiative corrections}
\label{sec_radcor}

To examine
the behaviour of the system under radiative corrections we adopt
an effective field theory (EFT) perspective.
  Between the pulses, the spectator field is exactly massless and enjoys a shift symmetry,
\begin{equation}
 \sigma \to \sigma+c\,.
\end{equation}
The shift symmetry is broken only at the pulse locations. This is different from a  spectator with a (small) bulk mass, where the symmetry is explicitly broken everywhere.
In fact, 
while for a bare, constant mass term we should expect quadratic corrections as \begin{equation}
 \Delta {\cal L}_{\rm bulk}
 \supset
 -\frac12 a^2\,\delta m_{\rm bulk}^2\,\sigma^2
\end{equation}
with $\delta m_{\rm bulk}^2\propto \Lambda^2$ ($\Lambda$ being the EFT cutoff), such
corrections are not allowed in our setup thanks to 
the  shift symmetry. 
Radiative effects which can break the shift symmetry only
occurs at the pulse positions, controlled by the discrete combination
in the function $J(\tau)$ of Eq.~\eqref{eq:finite-window-continuum}. After corrections
are included, we  expect a local effective action of the schematic form
\begin{equation}
 {\cal L}_{\rm eff}
 =
 -\frac12 a^2
 \left[J(\tau)+\delta J(\tau)\right]\sigma^2
 +
 \frac{c_\partial}{\Lambda^2}J(\tau)(\partial\sigma)^2
 +
 \frac{c_4}{\Lambda^4}J^2(\tau)\sigma^4
 +\cdots ,
 \label{eq:spurion_counterterms2}
\end{equation}
where again $\Lambda$ is a cut-off scale. 
The first term in the expression above
contributes to the pulse amplitudes, while the remaining terms deform the pulse
profile. In other words, this setup localizes radiative corrections at the pulse locations.

We now discuss  conditions to consider for making radiative corrections under control from an EFT perspective. 

\paragraph{Width of the localized source.}
A true delta function within the combination $J(\tau)$ is an idealization. A controlled EFT description should replace each pulse by a finite-width event,
\begin{equation}
 J(\tau)
 =
 \sum_j
 \frac{s_j}{\Delta\tau_j}
 F_j\!\left(
  \frac{\tau-\tau_j}{\Delta\tau_j}
 \right),
 \qquad
 \int dy\,F_j(y)=1\,.
 \label{eq:finite_width_source}
\end{equation}
The physical width is
$ \Delta t_j=a_j\Delta\tau_j
$, where $a_j=a(\tau_j)$. The EFT requires that the pulse is not shorter than the cutoff scale,
$ \Delta t_j\gtrsim \Lambda^{-1}$,
while the pulse should still be short compared with the Hubble time,
$ \Delta t_j\ll H^{-1}$. These two inequalities are compatible whenever
\begin{equation}
 H\ll \Lambda\,.
\end{equation}

\paragraph{Distance among pulses.}
For logarithmically spaced pulses, neighbouring events are separated by
\begin{equation}
 \Delta\tau_{{\rm sep},j}
 \simeq
 hT_j
\end{equation}
for $h\ll1$. Requiring the pulses to be individually resolved gives
$ \Delta\tau_j\ll hT_j$. 
Using $\Delta\tau_j\gtrsim 1/(a_j\Lambda)$ and, in de Sitter, $a_j=1/(HT_j)$, we find the  hierarchy
\begin{equation}
 \frac{H}{\Lambda}\ll h\ll1\,.
 \label{eq:hierarchy_h}
\end{equation}
The upper bound $h\ll1$ is the continuum condition, ensuring that the discrete tower 
can approximate the smooth Coulomb profile. The lower bound ensures that the finite-width pulses can fit between neighbouring events while remaining above the EFT cutoff.

\paragraph{Peak mass amplitude.}
There is also a constraint on the peak mass during a pulse. From \eqref{eq:finite_width_source}, the physical peak mass is roughly
\begin{equation}
 M_{{\rm pk},j}^2
 \sim
 \frac{s_j}{a_j^2\Delta\tau_j}
 =
 \frac{s_j}{a_j\Delta t_j}\,.
\end{equation}
The EFT requires
$ M_{{\rm pk},j}^2\ll \Lambda^2
$. 
If we express $\Delta t_j=\xi_j/\Lambda$ with $\xi_j\gtrsim1$, then
\begin{equation}
 M_{{\rm pk},j}^2
 \sim
 \frac{s_j\Lambda}{a_j\xi_j}\,.
\end{equation}
For the logarithmic tower, $s_j=h \kc$, and therefore
\begin{equation}
 \frac{s_j}{a_j}
 =
 h \kc H T_j\,.
\end{equation}
Thus a sufficient condition is
\begin{equation}
 h H \kc T_j\ll \xi_j\Lambda
\end{equation}
for all pulses included in the EFT. The strongest bound is usually at the earliest retained pulse, where $T_j$ is largest:
\begin{equation}
 h H \kc T_{\rm on}\ll \xi\Lambda\,.
 \label{eq:earliest_pulse_bound}
\end{equation}

\paragraph{Renormalization of the pulse amplitude}
Given the conditions above, we expect  that radiative corrections renormalize each pulse profile
\begin{equation}
 s_j\to s_j+\delta s_j\,,
\end{equation}
and generate higher localized operators.
 In principle, each pulse receives different corrections.
If the pulse contributions $\delta s_j/s_j$
can be set under control and made small, or if they are all the same -- for example exploiting the discrete
translational symmetry we invoked for motivating the tower
construction in Section \ref{subsec:logarithmic-tower} -- we can find conditions to make the setup radiative
stable. There are indications in favour of this perspective from a continuous Coulomb
limit of our framework, as we are going to discuss. 

\paragraph{The continuous limit} 
In the limit
of many pulses, so that we recover the effective Coulomb-potential
description of Section~\ref{subsec:exact-coulomb-filter}, we can
 assume the tree-level profile varies slowly in conformal time over the cutoff
scale, $| M_\sigma'/  M_\sigma|=1/(-\tau)\ll\Lambda$. Hence the expansion of $M_\sigma(\tau)$
in the leading correction of Eq.~\eqref{eq:spurion_counterterms2}
can be organized in powers of the small local ratio $M_\sigma/\Lambda^2$:
\begin{equation}
 \delta M_\sigma
 = c_0\,M_\sigma
 + c_1\,\frac{M_\sigma^2}{\Lambda}
 + c_2\,\frac{M_\sigma\,\partial_\tau^2 M_\sigma}{\Lambda^3}+\cdots .
 \label{eq:dJ-expansion}
\end{equation}
The leading term $\delta M_\sigma=c_0 M_\sigma$ is proportional to $M_\sigma$ itself:  it preserves the
$1/(-\tau)$ shape of the Coulomb potential, and only renormalizes the amplitude,
$ \kc\ \to\ \kc(1+c_0)$ in Eq.~\eqref{eq:ideal-coulomb-profile}. In this regime, then, we can argue that the structure 
of radiative corrections, under appropriate
assumptions on the behaviour
of the system well below the cutoff. Therefore, the Coulomb system can be set under
radiative  control from an EFT perspective.

\subsection{A massive-clock realization of the logarithmic pulse tower}
\label{subsec:massive-clock-realization}

The logarithmically
spaced pulse tower is a mechanism characterized by the universal
behaviour of $k^3$ growth from large towards small scales.  Many realizations of this universal properties can be
envisaged
 -- for example along the lines of the examples in Section \ref{subsec:curvature-pulses}.  In this section, as a proof of principle, 
 we develop a specific realization in terms of  an extra heavy spectator field as an
inflationary clock.  Massive fields oscillating during inflation are known to
record the evolution of the scale factor through features periodic in cosmic
time \cite{Chen:2014joa,Chen:2015lza} and, consequently, logarithmic in conformal time or momentum. We exploit the same property but for a
different purpose: the heavy clock periodically modulates the mass of the dark
matter spectator $\sigma$. We keep a low-energy, EFT perspective, while
commenting towards the end of the section on possible axion-type UV completions. 

\smallskip

There are two separate requirements to reproduce the pulse system introduced
above.  The clock must generate events equally spaced in physical time, which
automatically gives logarithmically spaced conformal times in de Sitter.  
But also, the integrated strength of each event in the canonical mode equation
must be independent of the event number.  The second condition is more difficult, 
because the amplitude of a freely oscillating massive field normally redshifts.

\smallskip

To make both properties manifest, define the number of e-folds along the
inflationary background by
\begin{equation}
 N(t)\equiv\ln\!\left(\frac{a(t)}{a_0}\right),
 \qquad
 N(\phi)\equiv
 -\int_{\phi_0}^{\phi}
 \frac{d\varphi}{\sqrt{2\epsilon(\varphi)}\,M_{\rm Pl}} .
 \label{eq:clock-N-definition}
\end{equation}
The second equality assumes that the inflaton decreases along its slow-roll
trajectory.  For approximately constant $H_{\rm inf}$ and slow-roll
parameter $\epsilon$ we have
\begin{equation}
 N\simeq H_{\rm inf}(t-t_0),
 \qquad
 N(\phi)\simeq
 -\frac{\phi-\phi_0}{\sqrt{2\epsilon}\,M_{\rm Pl}},
 \qquad
 T\equiv-\tau=T_0e^{-N},
 \label{eq:clock-N-relations}
\end{equation}
 We
consider the following effective potential for the inflaton $\phi$, a heavy
clock field $\xi$, and the spectator $\sigma$ on which we are interested to:
\begin{align}
 V(\phi,\xi,\sigma)
 ={}&V_{\rm inf}(\phi)
 +\frac12m_\xi^2
 \left[
  \xi-\Xi\cos\!\left(\nu N(\phi)+\theta_0\right)
 \right]^2
 \nonumber\\
 &+
 \frac12M_0^2e^{-N(\phi)}
 F_n\!\left(\frac{\xi}{\Xi}\right)\sigma^2,
 \qquad
 F_n(y)=y^{2n},
 \qquad n=1,2,\ldots .
 \label{eq:clock-potential}
\end{align}
with $V_{\rm inf}(\phi)$ the slow-roll inflationary potential, $m_\xi$ and
$M_0$ controlling the heavy clock $\xi$ and spectator $\sigma$ mass
respectively, $\nu$, $\Xi$ and $\theta_0$ tunable constants
with appropriate dimensions.
The clock is assumed to be  energetically subdominant, even though its time-dependent
minimum is driven by the slowly rolling inflaton.  Calling  $\omega$ the characteristic  clock frequency, when
\begin{equation}
H_{\rm inf}\ll  \omega\equiv \nu H_{\rm inf}  \ll 
 m_\xi ,
 \label{eq:clock-basic-hierarchy}
\end{equation}
the heavy field $\xi$ adiabatically follows the potential minimum:
\begin{equation}
 \xi(t)
 =
 \Xi\cos\!\left(\nu N(t)+\theta_0\right)
 +{\cal O}\!\left(
 \frac{\omega^2}{m_\xi^2},
 \frac{H_{\rm inf}\omega}{m_\xi^2}
 \right).
 \label{eq:clock-tracking-solution}
\end{equation}
For constant slow-roll $\epsilon$ parameter, the periodic minimum may equivalently be written as
$\Xi\cos(\phi/f+\theta_0)$, where
\begin{equation}
 f=\frac{\sqrt{2\epsilon}\,M_{\rm Pl}}{\nu},
 \qquad
 \omega=\frac{|\dot\phi|}{f}.
 \label{eq:clock-f-definition}
\end{equation}
The coupling in the second line of Eq.~\eqref{eq:clock-potential} then gives
the spectator a positive, periodically modulated mass,
\begin{equation}
 M_\sigma^2(t)
 =
 M_0^2e^{-N(t)}
 \cos^{2n}\!\left(\nu N(t)+\theta_0\right).
 \label{eq:clock-sigma-mass}
\end{equation}
The factor $e^{-N}$ is the slowly varying envelope required by the Coulomb
filter framework of Section \ref{subsec:logarithmic-tower}. Let
\begin{equation}
 C_n
 \equiv
 \frac{1}{\pi}\int_0^\pi d\theta\,\cos^{2n}\theta
 =
 \frac{(2n)!}{2^{2n}(n!)^2} .
 \label{eq:clock-Cn}
\end{equation}
Averaging Eq.~\eqref{eq:clock-sigma-mass} over one rapid clock period gives
\begin{equation}
 \overline{M_\sigma^2}(t)=C_nM_0^2e^{-N(t)}.
 \label{eq:clock-averaged-mass}
\end{equation}
It is useful to introduce the comoving scale
\begin{equation}
 {
 \kappa_c\equiv\frac{a_0C_nM_0^2}{H_{\rm inf}} \,,
 }
 \label{eq:clock-kc-definition}
\end{equation}
linking our setup with previous discussions. 
Using Eqs.~\eqref{eq:clock-N-relations} and
\eqref{eq:clock-kc-definition}, the complete modulated profile can be written
as
\begin{equation}
 a^2M_\sigma^2
 =
 \frac{\kc}{C_nT}
 \cos^{2n}\!\left[
  \nu\ln\!\left(\frac{T_0}{T}\right)+\theta_0
 \right],
 \label{eq:clock-modulated-Coulomb}
\end{equation}
whereas its coarse-grained limit results
$
 \overline{a^2M_\sigma^2}={\kappa_c}/{T}
 $.

\smallskip

\begin{figure}[t!]
 \centering
   \includegraphics[width=1\linewidth]{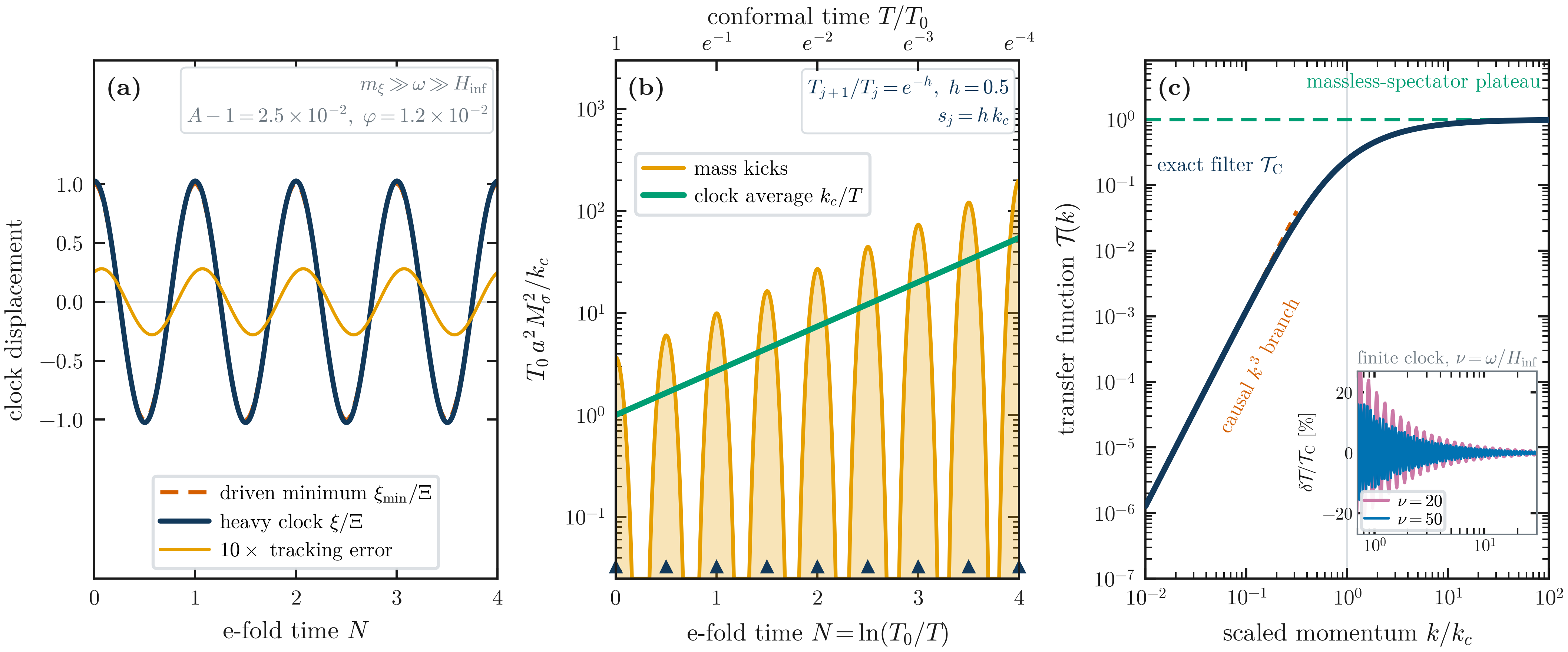}
 \caption{\small  Massive-clock realization of the Coulomb filter.
\textbf{(a)} A heavy spectator adiabatically tracks its driven minimum; the
tracking error (magnified $10\times$) is the small parameter
$\mathcal{O}(\omega^2/m_\xi^2)$. \textbf{(b)} The induced canonical mass
receives positive kicks equally spaced in $N$, hence geometrically spaced in
$T=-\tau$, with equal areas $s_j=h \kc$; their clock average is $\kc/T$.
\textbf{(c)} The exact transfer function, Eq.~\eqref{eq:coulomb-transfer},
interpolating between the causal $k^3$ branch and the massless plateau.
\emph{Inset:} finite-clock residuals obtained by integrating the modulated
profile of Eq.~\eqref{eq:clock-modulated-Coulomb}, log-periodic with
$\Delta\ln k\simeq\pi/\nu$.}
 \label{fig:clock}
\end{figure}

We can now recover directly both defining properties of the pulse tower.  The
maxima of Eq.~\eqref{eq:clock-sigma-mass} occur at
$ \nu N_j+\theta_0=j\pi$,
and are therefore separated by
$ \Delta N=\frac{\pi}{\nu}\equiv h$. 
Their conformal times consequently form the geometric sequence
\begin{equation}
 T_j=T_{\rm ref}e^{-jh},
 \qquad
 h=\frac{\pi}{\nu}
   =\frac{\pi H_{\rm inf}}{\omega}.
 \label{eq:clock-log-times}
\end{equation}
The conformal area of the $j$th event is
\begin{align}
 s_j
 &\equiv
 \int_{j\text{th event}}d\tau\,a^2M_\sigma^2
 =
 \int_{j\text{th event}}dt\,aM_\sigma^2
 \nonumber\\
 &=
 a_0M_0^2\int_{j\text{th event}}dt\,
 \cos^{2n}(\omega t+\theta_0)
 =
 \frac{\pi a_0C_nM_0^2}{\omega} .
 \label{eq:clock-pulse-area-intermediate}
\end{align}
These results imply
$ {s_j=h \kappa_c}$.
Hence the clock model generates exactly the logarithmic positions and constant
conformal areas assumed in the pulse construction.  Slow-roll evolution and
the finite response time of $\xi$ produce corrections suppressed by slow-roll
parameters and by powers of $\omega/m_\xi$.

\smallskip

For $n=1$, the realization is particularly simple.  Since $C_1=1/2$,
Eq.~\eqref{eq:clock-modulated-Coulomb} becomes
\begin{equation}
 a^2M_\sigma^2
 =
 \frac{\kc}{T}
 \left\{
 1+
 \cos\!\left[
  2\nu\ln\!\left(\frac{T_0}{T}\right)+2\theta_0
 \right]
 \right\}.
 \label{eq:clock-n1-profile}
\end{equation}
This is a rapidly modulated Coulomb profile.  The individual events are smooth
positive lobes rather than delta functions, but modes unable to resolve the
clock period respond only to the first term in Eq.~\eqref{eq:clock-n1-profile}.
For $n>1$, instead, the events become more sharply localized.  Near a maximum,
\begin{equation}
 \cos^{2n}(\omega\Delta t)
 \simeq e^{-n\omega^2\Delta t^2},
 \end{equation}
so that
\begin{equation}
 \Delta t_{\rm pulse}\sim\frac{1}{\omega\sqrt n},
 \qquad
 \frac{\Delta t_{\rm pulse}}{\Delta t_{\rm sep}}
 \sim\frac{1}{\pi\sqrt n}.
 \label{eq:clock-pulse-width}
\end{equation}
The choice $n=1$ is the minimal periodically modulated realization, while
$n>1$ gives a controlled approximation to a train of separated positive
kicks.

The Coulomb filter is recovered when the clock is fast compared with every
frequency resolved by the spectator modes during the relevant interval.  A
sufficient hierarchy is
\begin{equation}
\left(  H_{\rm inf},\quad \frac{k}{a},\quad M_\sigma \right)
 \ll\omega\ll m_\xi .
 \label{eq:clock-averaging-hierarchy}
\end{equation}
The first three inequalities ensure that the clock modulation can be
coarse-grained, while the last ensures that $\xi$ follows its moving minimum. These conditions make the clock construction an explicit realization
of the log-spaced pulses discussed in Section \ref{subsec:logarithmic-tower}. See Fig.~\ref{fig:clock} for a
representation of the system. We do not discuss  radiative 
corrections in this scenario. But we point out that higher order non-perturbative
corrections to axion realizations \cite{Parameswaran:2016qqq} can  modulate the axion potential introducing higher harmonics, leading
to oscillatory structures similar to the ones considered here. It would be interesting
to explore this avenue for an UV high energy condition of our setup.

\section{Filtered inflationary spectator dark matter}
\label{sec:filtered-spectator-dm}

The filter mechanism discussed so far in Sec.~\ref{sec:coulomb-filter-theory} is fully general. We now discuss one of its possible
cosmological applications.
We examine a {\it minimal} dark-matter scenario which  makes use of the filtered spectator
field. Our setup aims to make  minimalistic
assumptions on the spectator field dynamics during and after inflation,
 showing that we can obtain a working model of dark matter
with the  very few available parameters at our disposal. Dark matter is non-thermally 
produced by inflation: see e.g. \cite{Kolb:2023ydq} for a review. It is  associated with the  spectator $\sigma$ undergoing 
a standard misalignment mechanism \cite{Preskill:1982cy,Abbott:1982af,Dine:1982ah} -- see e.g. \cite{Marsh:2015xka} for a review -- although in our setup the spectator mass has a specific time dependence. 
 The inflaton
sector remains standard and generates the observed adiabatic curvature
perturbation. The Hubble scale during inflation is $H_{\rm inf}$.  In the first phases of cosmological expansion, we impose that the spectator field $\sigma$ is   dynamically subdominant and  uncorrelated with the inflaton,
up to indirect  interactions through gravity and through  sectors possibly responsible for its time-dependent
mass (recall the example in Section \ref{subsec:massive-clock-realization}).

\begin{figure}[t!]
 \centering
   \includegraphics[width=0.53\linewidth]{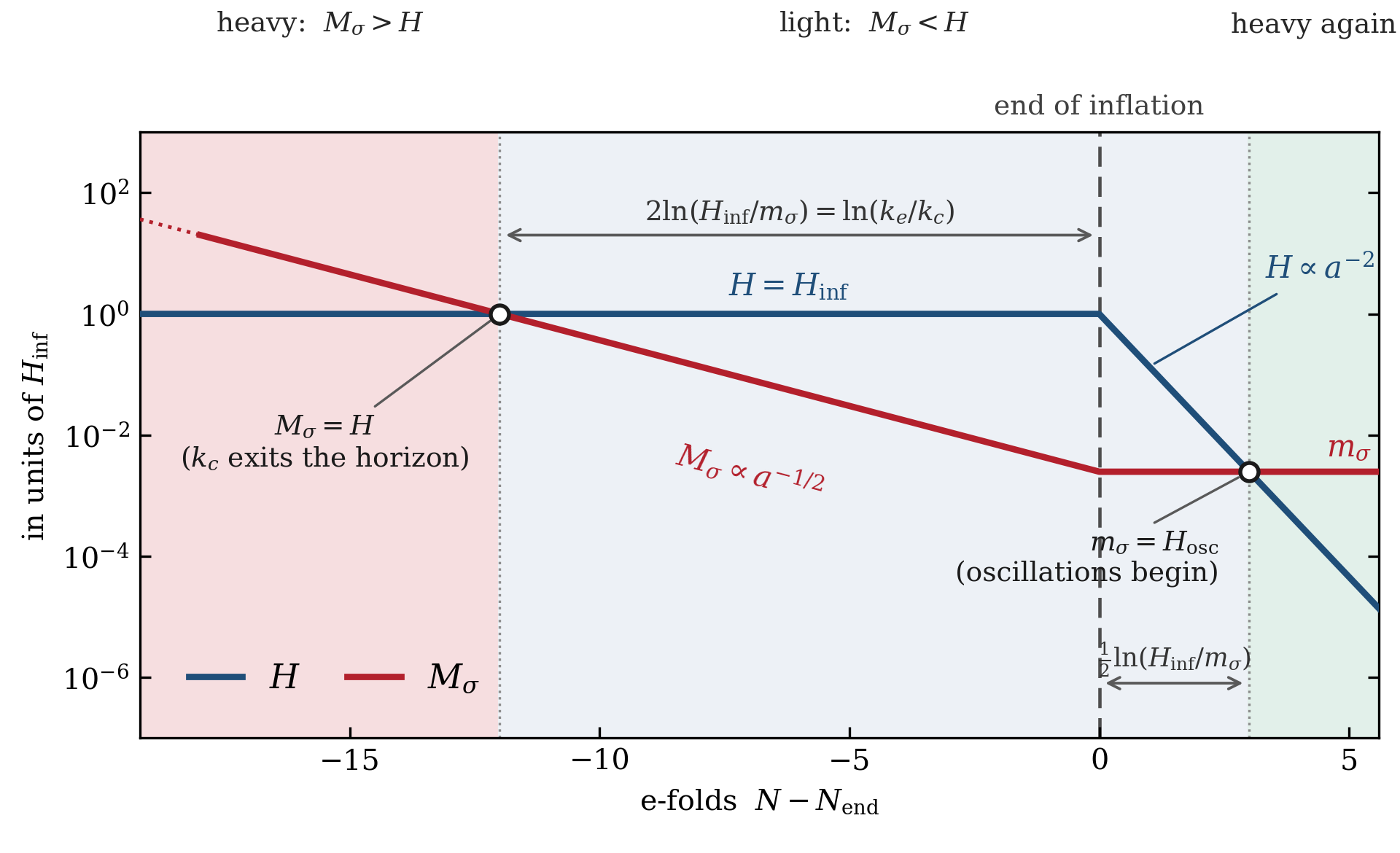}
      \includegraphics[width=0.46\linewidth]{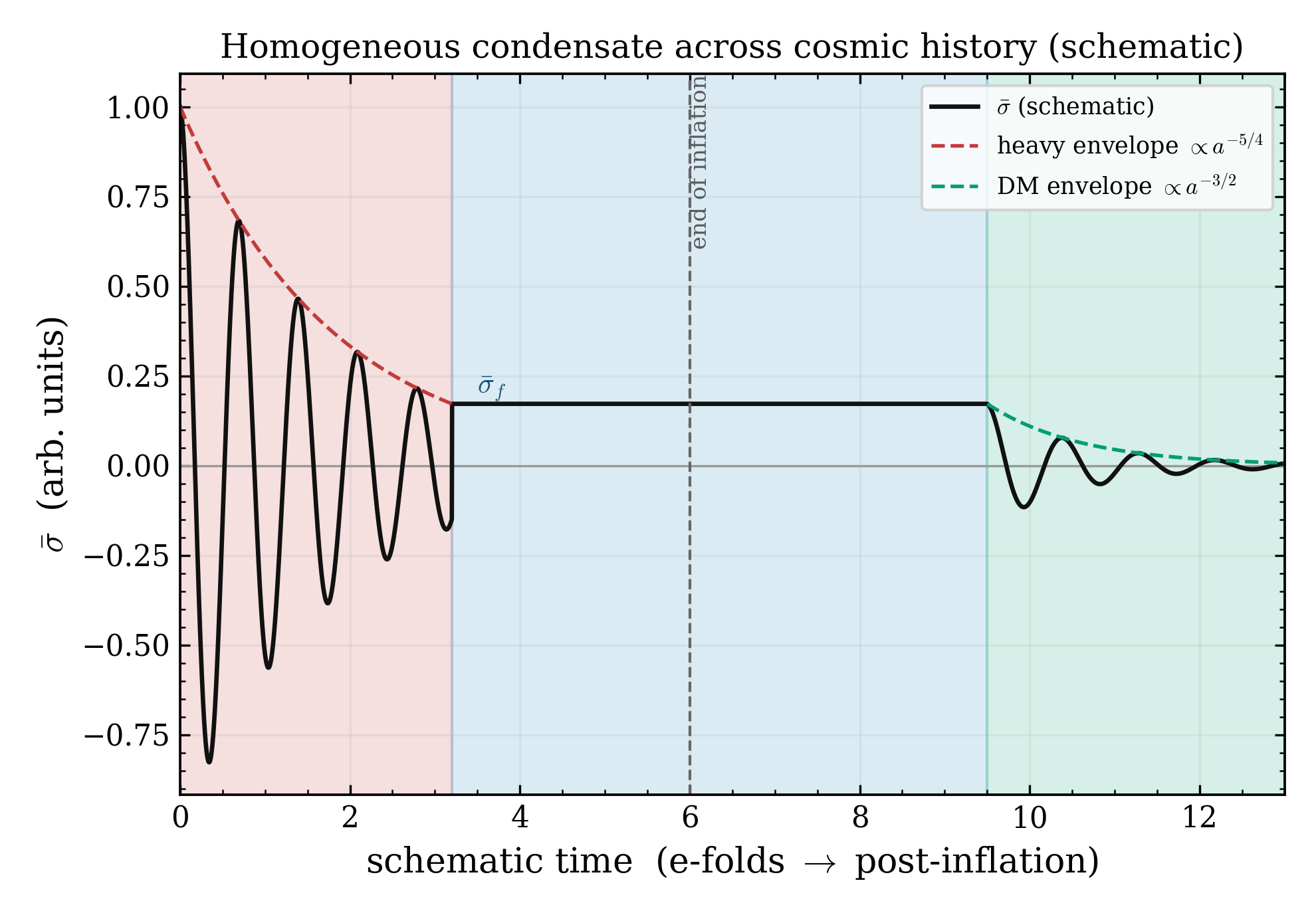}
  \caption{\small  Behaviour of the spectator field mass $M_\sigma$ ({\bf left panel})
 as well as the homogeneous spectator profile $\sigma$  ({\bf right panel})
 during cosmological history. Shaded colored regions correspond to phases
 in which the spectator mass can be larger or smaller than the Hubble scale,
 during or post inflation.
 }
 \label{fig:single-mass}
\end{figure}

\smallskip
We start  outlining the homogeneous evolution of the spectator through different cosmological epochs --  an analysis we develop more  quantitatively
in Section \ref{sub_thback}.  To build physical intuition, let us  examine Fig.~\ref{fig:single-mass}.    During inflation, in the Coulomb regime, the time-dependent spectator mass  
\begin{equation}
a^2 M_\sigma^2 = \kc/(-\tau)
\end{equation}
(recall Eq.~\eqref{eq:ideal-coulomb-profile}) is initially 
larger than the constant Hubble scale $H_{\rm inf}$. In this epoch the field oscillates. See the red shaded regions of Fig.~\ref{fig:single-mass}. At the
scale  $M_\sigma(\tau)=H_{\rm inf}$ -- when modes at turnover scale $\kappa_c$ leave the horizon -- the spectator field becomes lighter than the inflationary scale.
It stops oscillating and, by misalignment, it becomes a constant which we denote with $\bar \sigma_f$ (blue shaded regions of Fig.~\ref{fig:single-mass}). 
At the end of  inflation, the  process giving the time-dependent
spectator mass is no active any more.
 We make the hypothesis that the spectator freezes its mass at a value  $m_\sigma$. 
 After inflation, the Hubble parameter starts  decreasing, until a stage when $H=m_\sigma$. At this point,
the spectator starts oscillating again, leading to pressureless dark matter on average (green shaded regions of Fig.~\ref{fig:single-mass}).   

\smallskip
We can quantify the amount of dark matter
 in this scenario, relating it to the available parameters.
 Working at the homogeneous level, if the 
post-inflationary spectator oscillations
start during radiation domination  the standard misalignment estimate for the dark matter abundance is
(see e.g. \cite{Marsh:2015xka})
\begin{align}
  \Omega_\sigma h^2\simeq{}&0.12
  \left(\frac{\sigbar_f}{8.8\times10^{11}\,{\rm GeV}}\right)^2
  \left(\frac{\ms}{1\,{\rm eV}}\right)^{1/2}
  \left(\frac{g_{*,\rm osc}}{106.75}\right)^{3/4}
  \left(\frac{g_{*s,\rm osc}}{106.75}\right)^{-1} .
  \label{eq:misalignment_abundance}
\end{align}
Here $g_*$ and $g_{*s}$ denote the energy and entropy relativistic degrees of
freedom.  Setting $g_*=g_{*s}$ reproduces the often-used factor
$g_{*,\rm osc}^{-1/4}$.  The numerical coefficient carries the usual order-one
uncertainty associated with the precise oscillation criterion.
%
If $\sigma$ constitutes all of the dark matter,
\begin{equation}
  {
  \sigbar_{ f}\simeq
  8.8\times10^{11}\,{\rm GeV}
  \left(\frac{\ms}{1\,{\rm eV}}\right)^{-1/4}
  \left(\frac{g_{*,\rm osc}}{106.75}\right)^{-3/8}
  \left(\frac{g_{*s,\rm osc}}{106.75}\right)^{1/2} \,,
  }
  \label{eq:sigma_DM}
\end{equation}
Hence this formula directly relate $\sigbar_f$ with $m_\sigma$, the
constant spectator mass at the end of inflation. The latter
quantity can then be related with the turnover scale $\kappa_c$ of Eq.~\eqref{eq:coulomb-transfer}
by the following arguments. By definition, the final  mass is given by $ \ms^2=M_\sigma^2(\tau_e)
  =\hinf^2\kc(-\tau_e)$. Let us introduce now a CMB pivot scale $\ks$,
  corresponding to modes leaving the horizon early,  $N_*$ e-folds before the end of
inflation.  In exact de Sitter,
$  \ks=a_*\hinf= e^{-N_*}/(-\tau_e) \,\equiv\,e^{-N_*} k_e
$.
By combining these expressions,
we  introduce the ratio $ { R}$ between turnover and pivot scales: 
\begin{equation}
  {
   { R}\equiv\frac{\kc}{\ks}
  =e^{N_*}\left(\frac{\ms}{\hinf}\right)^2 \, \hskip0.5cm \Rightarrow
  \hskip0.5cm \ms=\hinf\sqrt{ { R}}\,e^{-N_*/2} \,.
  }
  \label{eq:key_kc_relation}
\end{equation}
Therefore, we can unambigously relate the spectator mass to the  spectrum turnover scale and the inflationary Hubble parameter, which are essentially
the only free parameters we are going to use. Our scenario producing primordial
dark matter hence results
very economical in terms of number of parameters involved.

\smallskip

We proceed with Section \ref{sub_thback}    discussing  theoretical
constraints that backreaction imposes on this spectator  dark matter scenario. 
Section \ref{sub_confex} explains how we set constraints from  large-scale
cosmological observations, developing
explicit examples of benchmark
models that satisfy theoretical and observational requirements.  Section \ref{sec:context}, then,
compares our setup with  other scenarios recently discussed
in the cosmology literature. 

\subsection{Theoretical conditions ensuring controlled backreaction}
\label{sub_thback}

We  spell out in more detail the theoretical characterization
of the  setup  outlined above,  focussing on 
constraints on our scenario should satisfy, due to  backreaction effects in the first epochs of cosmological evolution
 -- the red and blue bands in Fig. \ref{fig:single-mass}. 
\paragraph{Constraints from homogeneous backreaction}
\label{subsec:dm-backreaction}

The homogeneous condensate associated with $\sigma$ must remain a spectator throughout the entire interval
in which the mass profile is active during inflation. We now work in the Coulomb
dense limit of our description.
 Define the quantity 
\begin{equation}
  \chi(\tau)\equiv\kc(-\tau)
  =\frac{ M_\sigma^2(\tau)}{\hinf^2} .
  \label{eq:chi_definition}
\end{equation}
In the heavy regime $\chi\gg1$ during inflation (Fig.~\ref{fig:single-mass}, left panel, 
  red shaded region), the oscillation-averaged condensate energy~\footnote{We stress that these inflationary oscillations
are different from the post-inflationary, dark matter oscillations leading to a pressureless fluid.} is
approximately
\begin{equation}
  \rho_\sigma\simeq\frac12M_\sigma^2\sigbar_{\rm amp}^2,
\end{equation}
 We denote by
$\sigbar_{\rm amp}(\tau)$ the slowly evolving amplitude (envelope) of these
oscillations.
We introduce the quantity $B$ defining a homogeneous backreaction parameter:
%
%
\begin{equation}
  B(\tau)\equiv
  \frac{\rho_\sigma}{3\mpl^2\hinf^2}
  \simeq\frac{\chi(\tau)}{6}
  \left(\frac{\sigbar_{\rm amp}(\tau)}{\mpl}\right)^2 .
  \label{eq:backreaction_general}
\end{equation}
In the heavy regime $\chi\gg1$ the homogeneous field oscillates during inflation;
A controlled spectator approximation requires $B\ll1$.

\begin{figure}[t!]
  \centering
  \includegraphics[width=0.9\linewidth]{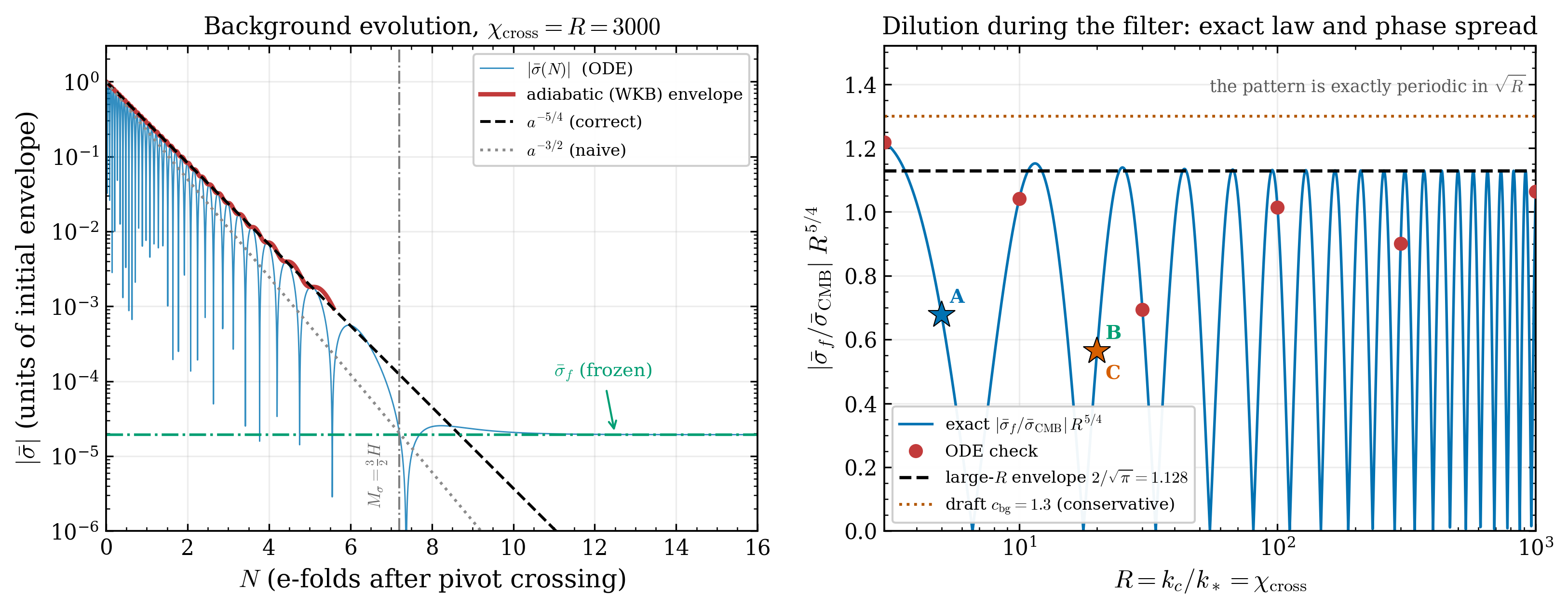}
  \caption{\small 
  Dilution of the homogeneous condensate while the filter operates.
  \emph{Left}: a representative trajectory, $R=3000$.  Once
  $M_\sigma>\tfrac32H$ the field oscillates and its adiabatic envelope decays
  as $a^{-5/4}$ (black dashed), not as the naive $a^{-3/2}$ (grey dotted): the
  growing mass makes the oscillator increasingly adiabatic, and the extra
  $a^{1/4}$ is the accumulated correction.  When $M_\sigma$ falls back below
  $\tfrac32H$ the field stops oscillating and freezes at $\sigbar_f$
  (green).  \emph{Right}: the frozen amplitude, rescaled by $R^{5/4}$.  The
  problem is exactly soluble --- with $\chi=Re^{-N}$ the substitution
  $\sigbar=e^{-3N/2}w$, $x=2\sqrt{\chi}$ reduces the equation to Bessel's of
  order three, so that
  $\sigbar_f/\sigbar_{\rm CMB}=(2/R)\left[3J_3(x_0)/x_0+J_3'(x_0)\right]$ with
  $x_0=2\sqrt{R}$, see Eq.~\eqref{eq:background_bessel_solution} --- and the blue curve is that
  expression, with direct integrations of the equation of motion (red circles)
  as a check.  The apparent scatter of the frozen amplitude is therefore not
  numerical noise but an exact oscillation, periodic in $\sqrt{R}$: it records
  the phase of the oscillation at freeze-out, and it passes through zero
  whenever the field freezes at a node.  The oscillation is bounded above by
  $2/\sqrt{\pi}\simeq1.128$ (black dashed). Stars mark the benchmark points  we will
  discuss in Section \ref{sub_confex}.
     }
  \label{fig:background_damping}
\end{figure}

Since we considered so far linearized fluctuations, 
the homogeneous evolution can be obtained from the same type
of Coulomb equation controlling the spectator fluctuations.  With
$T=-\tau$ and $u_{\rm bg}=a\sigbar$, one finds
\begin{equation}
  \frac{d^2u_{\rm bg}}{dT^2}
  +\left(\frac{\kc}{T}-\frac{2}{T^2}\right)u_{\rm bg}=0,
\end{equation}
whose solutions are expressed in terms of Bessel functions
\begin{equation}
  u_{\rm bg}(T)=\sqrt y\left[
  \mathcal A_3\, J_3\!\left(2\sqrt{\kc T}\right)
  +\mathcal B_3\, Y_3\!\left(2\sqrt{\kc T}\right)
  \right] .
  \label{eq:background_bessel_solution}
\end{equation}
The envelope of the solution scales
as
\begin{equation}
  {
  \sigbar_{\rm amp}\propto a^{-5/4},
  \qquad
  \rho_\sigma\propto a^{-7/2} .
  }
  \label{eq:background_scaling}
\end{equation}
with the scale factor. 
These scaling relations also follow from conservation of the adiabatic comoving particle
number: $n_\sigma\propto a^{-3}$ while
$M_\sigma\propto a^{-1/2}$. See Fig.~\ref{fig:background_damping}.

Let $\sigbar_f$ denote the approximately frozen late-inflationary displacement
after the mass has fallen below the Hubble scale during
the last stage of inflation.  Matching the heavy oscillatory
solution discussed above to this late inflationary frozen regime gives, up to a phase-dependent order-one factor
$c_{\rm bg}$, the relation
\begin{equation}
  \sigbar_{\rm amp}(\chi)
  \simeq c_{\rm bg}\,\sigbar_f\,\chi^{5/4},
  \qquad \chi = \kc T \gg1 .
  \label{eq:background_matching}
\end{equation}
Hence, towards the end of inflation, the backreaction parameter $B$ of Eq.~\eqref{eq:backreaction_general}  results
\begin{equation}
  {
  B(\chi)\simeq
  \frac{c_{\rm bg}^2\sigbar_f^2}{6\mpl^2}\,\chi^{7/2} 
  }  \hskip0.5cm \Rightarrow   \hskip0.5cm B_*\simeq
  \frac{c_{\rm bg}^2\sigbar_f^2}{6\mpl^2}\,{ R}^{7/2} .
\end{equation}
where the arrow in the previous formula matches  $B$
at pivot crossing, $-\ks\tau_*=1$ and at
$\chi_*=\kc/\ks={ R}$.

A small value for $B_*$
 is a useful diagnostic of controlled backreaction, but it is not automatically the strongest bound
 we need to impose in our setup.
In fact, since  $\rho_\sigma\propto a^{-7/2}$ during the stage of inflationary heavy mass, the largest
backreaction occurs at the \emph{earliest} time at which the Coulomb profile is
active.  If the physical filter starts $\Delta N_{\rm on,*}$ e-folds before pivot
crossing, then we can scale $ \chi_{\rm on}\,=\,{ R} \,e^{\Delta N_{\rm on,*}}$ (recall
the definition in Eq.~\eqref{eq:chi_definition}), and we  impose
\begin{equation}
  B_{\rm on}\simeq B_*e^{\frac72\Delta N_{\rm on,*}} 
  \ll 1\,.
  \label{eq:backreaction_onset}
\end{equation}
This condition ensures that the homogeneous evolution of the spectator system does not 
backreact on the inflationary phase.

\paragraph{Constraints from the variance of fluctuations}
\label{subsec:dm-validity-extensions}

Besides the homogeneous evolution of the spectator
field discussed above, we  need to  consider extra theoretical  constraints from the dynamics
of  its fluctuations, in order  to ensure we consistently work within a linear 
perturbative regime. 
 On every
scale used in the  analysis of isocurvature fluctuations we impose
\begin{equation}
  \epsilon_{\rm lin}(k)
  \equiv
  \frac{\sqrt{\mathcal P_{\delta\sigma}(k)}}{|\sigbar_f|}
  =\frac{\hinf}{2\pi|\sigbar_f|}\sqrt{\PT(k)}
  \ll1 .
  \label{eq:linear_density_condition}
\end{equation}
If this condition fails, the spectator-field contribution enters in a non-linear
regime, something that we wish to avoid (more on this later).


\smallskip

Nonetheless,
the mode-by-mode condition in Eq.~\eqref{eq:linear_density_condition} does not
by itself guarantee that the homogeneous condensate dominates the field
configuration.  In fact, even if every logarithmic interval has a small fluctuation
amplitude, the contributions from many fluctuation modes might accumulate.
But our analytical  control of the spectator
sector allows us to carry out an exact  quantitative analysis. 
  The relevant
quantity is the variance of the field after coarse graining on a physical
scale \(L\),
\begin{equation}
 \Sigma_L^2
 \equiv
 \left\langle\delta\sigma_L^2\right\rangle
 =
 \int_{k_{\rm IR}}^{k_{\rm UV}}
 d\ln k\,
 {\cal P}_{\delta\sigma}(k)\,
 W_L^2(k),
 \label{eq:coarse-grained-variance}
\end{equation}
where \(W_L(k)\) is a smoothing window.  The infrared cutoff specifies the
largest region included in the local background, while the ultraviolet
cutoff is the largest mode that is both produced during inflation, and
retained in the coarse-grained condensate at the epoch of interest.

For our Coulomb spectrum given in Eq.~\eqref{eq:coulomb-transfer} it is useful to define an effective number of contributing logarithmic
intervals,
\begin{equation}
 {\cal N}_{\rm eff}(L)
 \equiv
 \int_{k_{\rm IR}}^{k_{\rm UV}}
 d\ln k\,
 {\cal T}_{\rm C}(k)\,
 W_L^2(k).
 \label{eq:Neff-definition}
\end{equation}
The infrared part of the integral is automatically convergent because
\({\cal T}_{\rm C}\propto k^3\).  If all retained modes lie below the
turnover, \(k_{\rm UV}\ll \kc\), then
\begin{equation}
 {\cal N}_{\rm eff}
 \simeq
 \frac{4}{3\pi}
 \left(\frac{k_{\rm UV}}{\kc}\right)^3,
 \label{eq:Neff-filtered}
\end{equation}
up to corrections depending on the smoothing window.  By contrast, if the
integration range extends well into the ultraviolet plateau,
\(k_{\rm IR}\ll \kc\ll k_{\rm UV}\), one obtains
\begin{equation}
 {\cal N}_{\rm eff}
 \simeq
 \ln\!\left(\frac{k_{\rm UV}}{\kc}\right)
 +C_{\rm C},
 \qquad
 C_{\rm C}\simeq-0.95,
 \label{eq:Neff-plateau}
\end{equation}
for a sharp momentum-space window.  The order-one constant \(C_{\rm C}\)
originates  from the smooth Coulomb turnover associated with the scale
$\kc$ in Eq.~\eqref{eq:coulomb-transfer}.  It is interesting
that our smooth, analytical expression
for the filter  allows us to attain  such analytical control on ${\cal N}_{\rm eff}$.
The important result is the
logarithmic accumulation of the unsuppressed plateau modes.

A useful upper estimate for the fluctuation contributions is  obtained by including all modes that have crossed
the horizon by the end of inflation, \(k_{\rm UV}=k_e\equiv a_eH_{\rm inf}\).
In the minimal  completion,
$ \kc={a_e m_\sigma^2}/{H_{\rm inf}},
$ and therefore, when \(m_\sigma<H_{\rm inf}\),
\begin{equation}
 \ln\!\left({k_e}/{\kc}\right)
 =
 2\ln\!\left({H_{\rm inf}}/{m_\sigma}\right).
 \label{eq:plateau-duration-final-mass}
\end{equation}
Thus the ultraviolet plateau can span many e-folds even when the
fluctuation amplitude in each individual interval is perturbative.  Notice that a
coarser smoothing scale, or a requirement that the modes remain coherent
until the onset of oscillations, lowers \(k_{\rm UV}\), and  therefore it reduces this
estimate.

Hence,
the homogeneous-misalignment conditions  require
\begin{equation}
 {
 \epsilon_{\rm var}^2(L)
 \equiv
 \frac{\Sigma_L^2}{\bar\sigma_f^2}
 =
 \left(\frac{H_{\rm inf}}
 {2\pi\bar\sigma_f}\right)^2
 {\cal N}_{\rm eff}(L)
 \ll1 .
 }
 \label{eq:integrated-linearity-condition}
\end{equation}
This condition is stronger than the mode-by-mode criterion whenever
\({\cal N}_{\rm eff}\gg1\).  It ensures that the abundance is dominated by
the homogeneous displacement,
\begin{equation}
 \left\langle\rho_\sigma\right\rangle_{\rm osc}
 \simeq
 \frac12m_\sigma^2
 \left(
   \bar\sigma_f^2+\Sigma_L^2
 \right)
 \simeq
 \frac12m_\sigma^2\bar\sigma_f^2,
 \label{eq:abundance-with-variance}
\end{equation}
and that the density perturbation may be linearized as
\begin{equation}
 \delta\rho_\sigma
 \simeq
 m_\sigma^2\bar\sigma_f\,\delta\sigma.
\end{equation}

If Eq.~\eqref{eq:integrated-linearity-condition} is not satisfied, the
mechanism need not be inconsistent, but it leaves the condensate-dominated
regime studied here.  The mean abundance would receive an appreciable
contribution from the inflationary fluctuations, while
\begin{equation}
 \delta\rho_\sigma
 =
 m_\sigma^2\bar\sigma_f\,\delta\sigma
 +
 \frac12m_\sigma^2
 \left(
   \delta\sigma^2-\Sigma_L^2
 \right)
 \label{eq:density-linear-quadratic}
\end{equation}
contains an intrinsically non-Gaussian quadratic component.  Determining the
resulting abundance and isocurvature statistics would require keeping the
full finite-band spectrum and its post-inflationary transfer -- something
that we do not do here. In what comes next we therefore
impose Eq.~\eqref{eq:integrated-linearity-condition} as an additional
consistency condition of the minimal homogeneous-condensate realization,
without attempting a separate fluctuation-dominated analysis.

\subsection{Confronting the model with observations}
\label{sub_confex}

After discussing the  framework 
defining our dark matter setup at the homogeneous level, and analyzing theoretical 
consistency conditions associated with  requirements
of negligible backreaction, it is time to  examine
observational constraints from current  limits on isocurvature
fluctuation to large-scale cosmological observations. As we shall
learn, the filtering mechanism discussed in Section \ref{sec:coulomb-filter-theory}  becomes
essential for making our scenario viable. 

\paragraph{Linear isocurvature spectrum and  CMB physics.}
\label{subsec:dm-cmb-isocurvature}
As discussed above, 
for realizing a   spectator condensate dark matter in a quadratic potential  we require
 a small size of fluctuations: 
 $|\delta\sigma|\ll|\sigbar_f|$. Hence, we can 
 write
 at linear order
\begin{equation}
  \frac{\delta\rho_\sigma}{\rho_\sigma}
  \simeq2\frac{\delta\sigma}{\sigbar_f} .
\end{equation}
If $\sigma$ contributes a fraction $f_\sigma$ of the total cold dark matter, its
contribution to the total cold dark matter isocurvature contributions
(from now on abbreviated to CDI) is therefore
\begin{equation}
  S_{\rm cdm}\simeq2f_\sigma\frac{\delta\sigma}{\sigbar_f} .
\end{equation}
where
$S_{\rm cdm}\equiv\delta\rho_{\rm cdm}/\rho_{\rm cdm}-\,
(3 \delta\rho_\gamma)/(4 \rho_\gamma)$, evaluated on large scales in the radiation
era.

Using the Coulomb-filter spectrum obtained in Sec.~\ref{sec:coulomb-filter-theory}, and imposing the fractional abundance relation
\eqref{eq:misalignment_abundance}, we  obtain the isocurvature spectrum
\begin{equation}
  {
  \Ps(k)=f_\sigma
  \left(\frac{\hinf}{\pi\sigbar_{\rm DM}}\right)^2{\cal T}_c(k) .
  }
  \label{eq:isocurvature_power_fraction}
\end{equation}
Thus a subdominant condensate relaxes the observable isocurvature power linearly in
$f_\sigma$, rather than quadratically. We define the primordial isocurvature fraction at the CMB pivot scale by
the relation
\begin{equation}
  \beta_{\rm iso}(\ks)
  \equiv
  \frac{\Ps(\ks)}{\mathcal P_{\cal R}(\ks)+\Ps(\ks)} .
\end{equation}
Imposing $\beta_{\rm iso}\ll1$,
\begin{equation}
  \Ps(\ks)\lesssim
  \frac{\beta_{\rm iso}^{\rm max}}{1-\beta_{\rm iso}^{\rm max}}A_s
  \simeq\beta_{\rm iso}^{\rm max}A_s .
  \label{eq:CMB_condition}
\end{equation}
Here $A_s\equiv{\cal P}_{\cal R}(\ks)\simeq2.1\times10^{-9}$ is the measured
adiabatic amplitude at the pivot. 
Planck gives a representative $95\%$ bound
$\beta_{\rm iso}<0.038$ for an uncorrelated, nearly scale-invariant CDI template
at $k=0.05\,{\rm Mpc}^{-1}$~\cite{Planck:2018jri}.  Our spectrum is not scale
invariant, so this number is not an exact likelihood bound which can be directly applied
to  the Coulomb model.
It is nevertheless a useful reference.  We choose 
$\beta_{\rm iso}^{\rm max}=10^{-2}$ obtaining  a  conservative 
criterion.  

The comparison with bounds \cite{Planck:2018jri}  is, however, conservative in a precise pointwise sense -- as we can analytically demonstrate using the exact form of our transfer function \eqref{eq:coulomb-transfer}. Since
$1-e^{-\pi u}\le\min(1,\pi u)$ for $u=\kc/k>0$, the exact transfer function
obeys
\begin{equation}
  \mathcal T_{\rm C}(k)
  \;\le\;
  \min\!\left[\frac{4}{\pi}\left(\frac{k}{\kc}\right)^{3},\,1\right]
  \qquad\text{for all }k ,
  \label{eq:pointwise-bpl-bound}
\end{equation}
with equality only in the two asymptotic limits. The smooth Coulomb spectrum
therefore carries \emph{strictly less} power than the asymptotically matched
broken power law at every scale; near the break the deficit is a factor of a
few (e.g.\ $\mathcal T_{\rm C}(\kc)\simeq0.24$). Any upper limit derived for
the sharp template with the identification
\eqref{eq:benchmark-asymptotic-map} thus over-covers the primordial power of
the smooth spectrum scale by scale, and points allowed for the template are
expected to be allowed  for the Coulomb shape.

\paragraph{Constraints from small-scale structure.}
\label{subsec:dm-small-scale}

The filter suppresses isocurvature rather than removing it: we now need 
examine what happens at scales smaller than the CMB. 
After defining 
 the
ultraviolet plateau amplitude
\begin{equation}
 A_{\rm C}
 \equiv
 f_\sigma
 \left(\frac{H_{\rm inf}}{\pi\bar\sigma_{\rm DM}}\right)^2,
 \label{eq:benchmark-AC}
\end{equation}
we can write
  the asymptotic isocurvature 
spectrum is
\begin{equation}
 {\cal P}_S(k)
 \simeq
 A_{\rm C}
 \begin{cases}
 \displaystyle \frac{4}{\pi}\left(\frac{k}{k_0}\right)^3,
 & k\ll k_0,\\[7pt]
 1,&k\gg \kc.
 \end{cases}
 \label{eq:benchmark-coulomb-asymptotics}
\end{equation}

Consequently, a benchmark that is safe at the CMB pivot can still produce a
substantial CDI component on galaxy and sub-galaxy scales.
Here we need to be careful since
a pointwise comparison such as
$\Ps(k_{\rm Ly\alpha})\lesssim A_s$ is not a statistically meaningful
Lyman-$\alpha$ bound.  Adiabatic and CDI initial conditions have different
transfer functions, and the observable flux power depends on the amplitude and
slope of the evolved matter spectrum as well as on the thermal and astrophysical
history of the intergalactic medium.  The primordial spectrum should  ideally be
propagated with a Boltzmann solver and compared with an appropriate likelihood.

\smallskip

However, even without carrying on sophisticated
analysis at this stage,  we can exploit
recent works which are particularly relevant for us, because they consider precisely the
spectral class we naturally  obtain in our setup.  \cite{Buckley:2025zgh} constrains a
broken CDI spectrum proportional to $k^3$ below a turnover and constant above it,
using CMB, BAO, Lyman-$\alpha$ forest and CMB spectral-distortion data over
approximately $10^{-4}\lesssim k/{\rm Mpc}^{-1}\lesssim10^4$.  The Coulomb
spectrum has the same two asymptotic slopes -- but a smooth analytic turnover and an
order-one difference in the definition of the break scale. See also \cite{Kasuya:2009up,Chung:2017uzc} for related constructions.
  The broken-power-law
limits of Ref.~\cite{Buckley:2025zgh} are therefore the closest existing no-new-analysis comparison.

Complementarily, \cite{Co:2026klr} uses HST and JWST UV
luminosity functions to constrain general CDI spectra on
$k\simeq0.5$--$10\,{\rm Mpc}^{-1}$ and construct pointwise $68\%$ and $95\%$
credible envelopes.  These data directly probe the scales on which the Coulomb
spectrum can rise toward its plateau.  The published UV-luminosity and
Lyman-$\alpha$ results are broadly consistent over their overlapping range. But
their interpretation still depends on the assumed spectral family and on the
mapping to the matter spectrum.

Accordingly,  we adopt the following limited claim.  The CMB condition
\eqref{eq:CMB_condition} defines the large-scale design requirement.  Smaller-scale
data, as discussed above, provide an additional and potentially stronger test whenever the turnover
lies at or below the range $k\sim0.5$--$10\,{\rm Mpc}^{-1}$.  
 We discuss in the next section how we handle those data, and provide consistent benchmark scenarios for spectator dark matter within our framework.

\paragraph{Representative benchmarks satisfying  our constraints.}
\label{subsec:dm-benchmarks}

We now present concrete parameter choices for the dark-matter condensate realization.  Namely, we identify examples for which the Coulomb filter is
actually needed to satisfy the large-scale isocurvature condition.
We impose  homogeneous backreaction and  integrated fluctuation variance 
constraints at
the same points.   We  compare the resulting smooth spectra with
published constraints on blue cold-dark-matter isocurvature. For simplicity,
 we assume that the ideal Coulomb spectrum `dense' limit of Eq.~\eqref{eq:coulomb-transfer} applies
 through the entire evolution of the system. Correspondingly,
we identify three
suitable benchmark points, collected in Table \ref{tab:coulomb-benchmarks}.

\smallskip

To compare with data, we choose for definiteness 
$ N_*=55,
 \,
 k_*=0.05\ {\rm Mpc}^{-1},
 \,
 A_s=2.1\times10^{-9},
 \label{eq:benchmark-fixed-inputs}
$ and we use the abundance normalization in
Eq.~\eqref{eq:sigma_DM}.
 We match   the ultraviolet plateau and the infrared coefficients of Eq.~\eqref{eq:benchmark-coulomb-asymptotics},  obtaining 
\begin{equation}
 {
 A_{\rm iso}=A_{\rm C},
 \qquad
 k_0=\left(\frac{\pi}{4}\right)^{1/3}\kc
 \simeq0.923\,\kc .}
 \label{eq:benchmark-asymptotic-map}
\end{equation}
The published CDI limits are displayed in terms of
$(\Omega_i/\Omega_m)^2 A_{\rm iso}$.  We therefore use
$ q_c\equiv {\Omega_c}/{\Omega_m}\simeq0.84
$
and compare $q_c^2A_{\rm C}$ with the CDI curve. 
We stress again that equation
\eqref{eq:benchmark-asymptotic-map} is not an exact identification of the two
likelihoods. It preserves the infrared and ultraviolet limits, but {\it not} the
order-one shape of the turnover.

\begin{table}[t]
 \centering
 \resizebox{\linewidth}{!}{%
 \begin{tabular}{c c c c c c c c c}
  \toprule
  Benchmark
  & $H_{\rm inf}$ [GeV]
  & $m_\sigma$ [GeV]
  & ${\cal R}$
  & $\kc$ [$\mathrm{Mpc}^{-1}$]
  & $q_c^2A_{\rm C}$
  & ${\cal P}_S(k_*)/A_s$
  & $B_{\rm on}$
  & $\epsilon_{\rm var}^2$
  \\
  \midrule
  A
  & $3.0\times10^6$
  & $7.65\times10^{-6}$
  & $5.00$
  & $0.250$
  & $7.27\times10^{-11}$
  & $4.30\times10^{-4}$
  & $1.04\times10^{-7}$
  & $1.35\times10^{-9}$
  \\
  B
  & $5.0\times10^6$
  & $2.55\times10^{-5}$
  & $20.0$
  & $1.00$
  & $3.69\times10^{-10}$
  & $3.91\times10^{-5}$
  & $9.37\times10^{-4}$
  & $6.67\times10^{-9}$
  \\
  C
  & $7.0\times10^6$
  & $3.57\times10^{-5}$
  & $20.0$
  & $1.00$
  & $8.55\times10^{-10}$
  & $9.07\times10^{-5}$
  & $7.92\times10^{-4}$
  & $1.55\times10^{-8}$
  \\
  \bottomrule
 \end{tabular}}
 \caption{\small Representative benchmarks for the minimal all-dark-matter
 condensate realization.  The quantity $B_{\rm on}$ uses the finite-onset
 prescription as discussed around Eq.~\eqref{eq:early-onset-condition} and the
 representative matching coefficient $c_{\rm bg}=1.3$.  The variance ratio is
 $\epsilon_{\rm var}^2=\Sigma_L^2/\bar\sigma_f^2$, evaluated within the
  ultraviolet range described in the text.}
 \label{tab:coulomb-benchmarks}
\end{table}

\begin{figure}[t]
 \centering
 \includegraphics[width=0.8\linewidth]
 {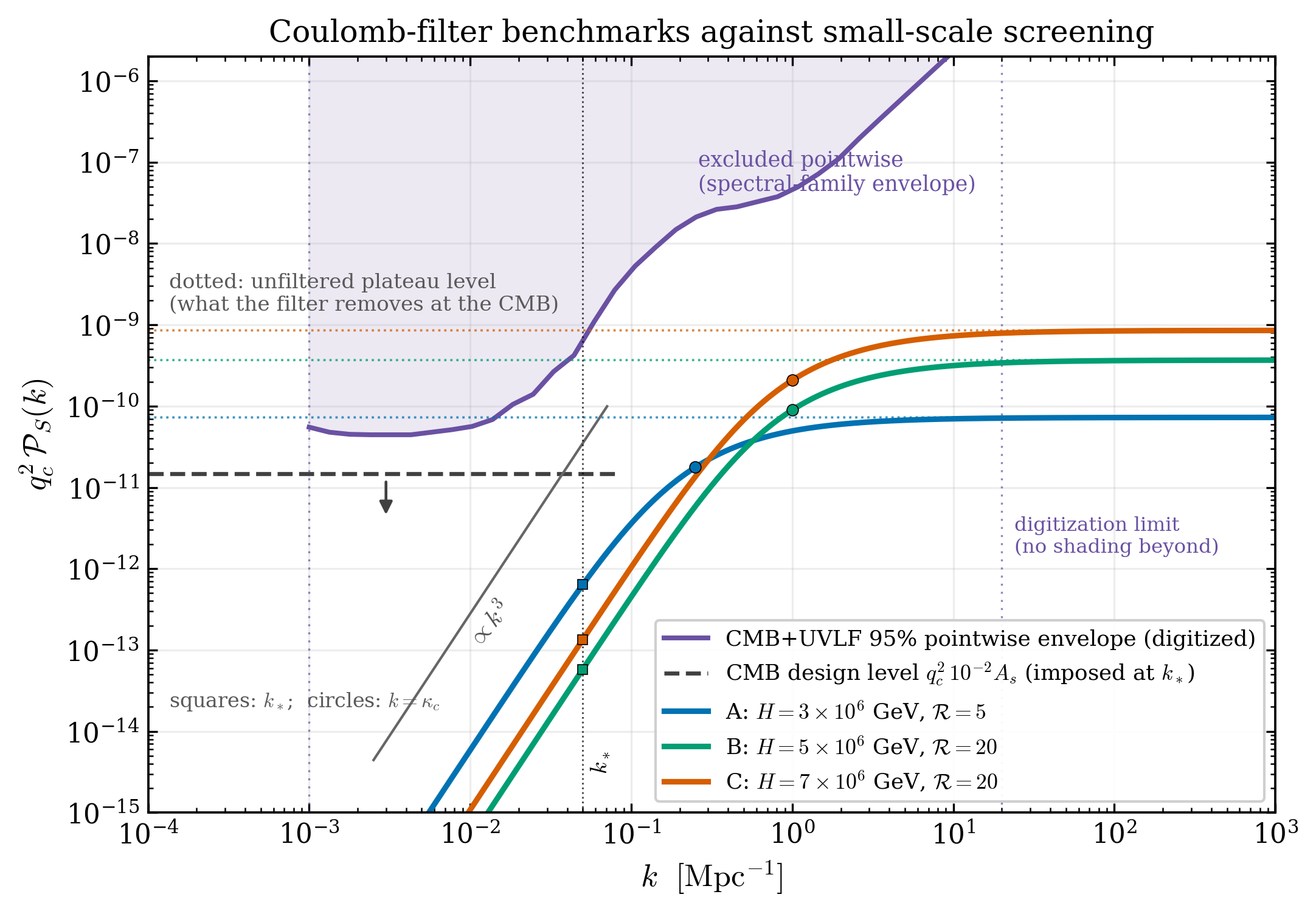}
 \caption{\small
 Exact weighted CDI spectra $q_c^2\Ps(k)$ for the benchmarks of
  Table~\ref{tab:coulomb-benchmarks} (solid; circles mark $k=\kc$).  Dotted
  horizontal lines give the corresponding \emph{unfiltered} plateau levels
  $q_c^2A_{\rm C}$, which exceed the CMB design level by factors of $5$ to $60$:
  the filter is doing genuine work at all three points.  The black dashed line is
  the design level $q_c^2\,10^{-2}A_s$ of Eq.~\eqref{eq:CMB_condition}, imposed at
  the pivot and drawn over the range $k\lesssim0.2\,\mathrm{Mpc}^{-1}$ on which a
  nearly scale-invariant CDI template is constrained.  Shaded: the complement of
  the digitized $95\%$ pointwise envelope of the combined CMB$+$UV luminosity
  function analysis of Ref.~\cite{Co:2026klr}, which bounds the largest power
  allowed at each scale across a family of spectral shapes.  Entering the shaded
  area at any scale disfavours a spectrum; staying below it everywhere is a
  screening test passed, not a likelihood evaluation (see text).  The shading
  terminates at $k=20\,\mathrm{Mpc}^{-1}$ because the published envelope is not
  quoted beyond that scale.  The thin gray line indicates the $k^3$ infrared
  slope.
   }
 \label{fig:benchmark-exact-spectraA}
\end{figure}

Figure~\ref{fig:benchmark-exact-spectraA} shows the exact weighted isocurvature spectra
$q_c^2\Ps(k)$ of the three benchmarks, together with two reference levels of 
different status.  The first is the large-scale design requirement
\eqref{eq:CMB_condition}, drawn as a black dashed line at $q_c^2\,10^{-2}A_s$.
It is a criterion {we} impose at the pivot, indicated over the range
$k\lesssim0.2\,\mathrm{Mpc}^{-1}$ on which a nearly scale-invariant CDI template
is  constrained.  The second is the shaded region, which encodes
observational information and is discussed below.  Before turning to it, note the
role of the dotted horizontal lines: they mark the \emph{unfiltered} plateau
levels $q_c^2A_{\rm C}$, that is, the isocurvature amplitude each benchmark would
carry if the spectator mass never switched on during inflation.  All three exceed
the design level, by factors ranging from about $5$ for benchmark A to about $60$
for benchmark C.  The vertical distance between each dotted line and the
corresponding solid curve at $k\simeq\ks$ is therefore a direct measure of the
work done by our Coulomb filter, and it establishes that these are genuine
filtering benchmarks rather than parameter choices for which an ordinary massless
spectator would already have been safe.

Let us now explain the construction of
the shaded area. It is the complement of the $95\%$ pointwise credible envelope on
CDI power obtained from the combined CMB and HST/JWST UV luminosity function
analysis of Ref.~\cite{Co:2026klr}, which we have digitized over the range
$10^{-3}\lesssim k/\mathrm{Mpc}^{-1}\lesssim20$ on which it is quoted.  At each wavenumber the envelope records the largest isocurvature
power allowed at \emph{that} scale when one scans over a family of spectral
shapes. Distinct points along the curve are in general saturated by distinct
members of the family.  It is then an upper hull, not a simultaneous
credible region for any single spectrum.
If a candidate spectrum enters the shaded area at \emph{any} scale, it
carries more power there than any member of the family was permitted to carry,
and it is  disfavoured.  Entering the shading is thus a sufficient
condition for rejection, and this is the sense in which the region is an
exclusion region.
 The three benchmarks are outside it  by a wide
margin.

Besides observational constraints, our benchmark
points in Table \ref{tab:coulomb-benchmarks} satisfy the theoretical requirements of controllable
backreaction, as discussed in  Section \ref{sub_thback} (to which we refer to for the definitions of quantities involved).  Let us explain
why. Writing 
\begin{equation}
 { R}\equiv\frac{\kappa_c}{k_*},
 \qquad
 x_{\rm on}\equiv k_*T_{\rm on},
\end{equation}
we choose the reference prescription
$x_{\rm on}=10\,R$, which satisfies $x_{\rm on}\gg1$ for all three benchmarks.
As discussed below Eq.~\eqref{eq:early-onset-condition}, the second
condition in Eq.~\eqref{eq:early-onset-condition} is not imposed: it cannot be
satisfied on CMB scales at fixed $\chi_{\rm on}$, and is replaced by the
requirement~\eqref{eq:onset-adiabatic-condition} that the profile be switched on
over $\mathcal{O}(1)$ e-folds, under which
Eq.~\eqref{eq:coulomb-transfer} holds to better than a percent across the whole
observational window.  The effective
spectator mass at onset is then
%
\begin{equation}
 \chi_{\rm on}=\kc T_{\rm on}
 ={ R}\,x_{\rm on}=10{ R}^2.
 \label{eq:benchmark-chi-on}
\end{equation}
Using Eq.~\eqref{eq:background_matching}, we estimate
\begin{equation}
 B_{\rm on}
 \simeq
 \frac{c_{\rm bg}^2\bar\sigma_f^2}{6M_{\rm Pl}^2}
 \chi_{\rm on}^{7/2},
 \qquad c_{\rm bg}=1.3,
 \label{eq:benchmark-Bon}
\end{equation}
where $c_{\rm bg}=1.3$ is a representative 
value~\footnote{Coming back to 
 to   Fig.~\ref{fig:background_damping}, we represent 
   $c_{\rm bg}=1.3$ as an
  orange dotted line
    above the exact envelope of the curve:  we regard this value as
   conservative for every $R$, since it overestimate  the surviving
  condensate amplitude and hence the backreaction.}
 of the phase-dependent
matching factor discussed below Eq.~\eqref{eq:background_matching}.
   The dependence of $B_{\rm on}$ on the onset prescription
should  be kept in mind: it is not a universal function of $H_{\rm inf}$ and
$m_\sigma$ alone.

\smallskip

Table~\ref{tab:coulomb-benchmarks}, as anticipated, gives three screened examples
of parameter
choices that satisfy {\it all} our requirements.  The exact
Coulomb function, rather than its infrared approximation, is used at the
pivot.  The finite-band variance is evaluated by retaining all modes that
leave the horizon by the end of inflation, as discussed around Eq.~\eqref{eq:integrated-linearity-condition},
 which is a conservative choice for the homogeneous-condensate test. Collecting
 all the information, we can determine the region of interesting parameter space
 within the plane spanned by the parameters $H_{\rm inf}$ and $m_\sigma$.

\begin{figure}[t!]
 \centering
  \includegraphics[width=0.8\linewidth]{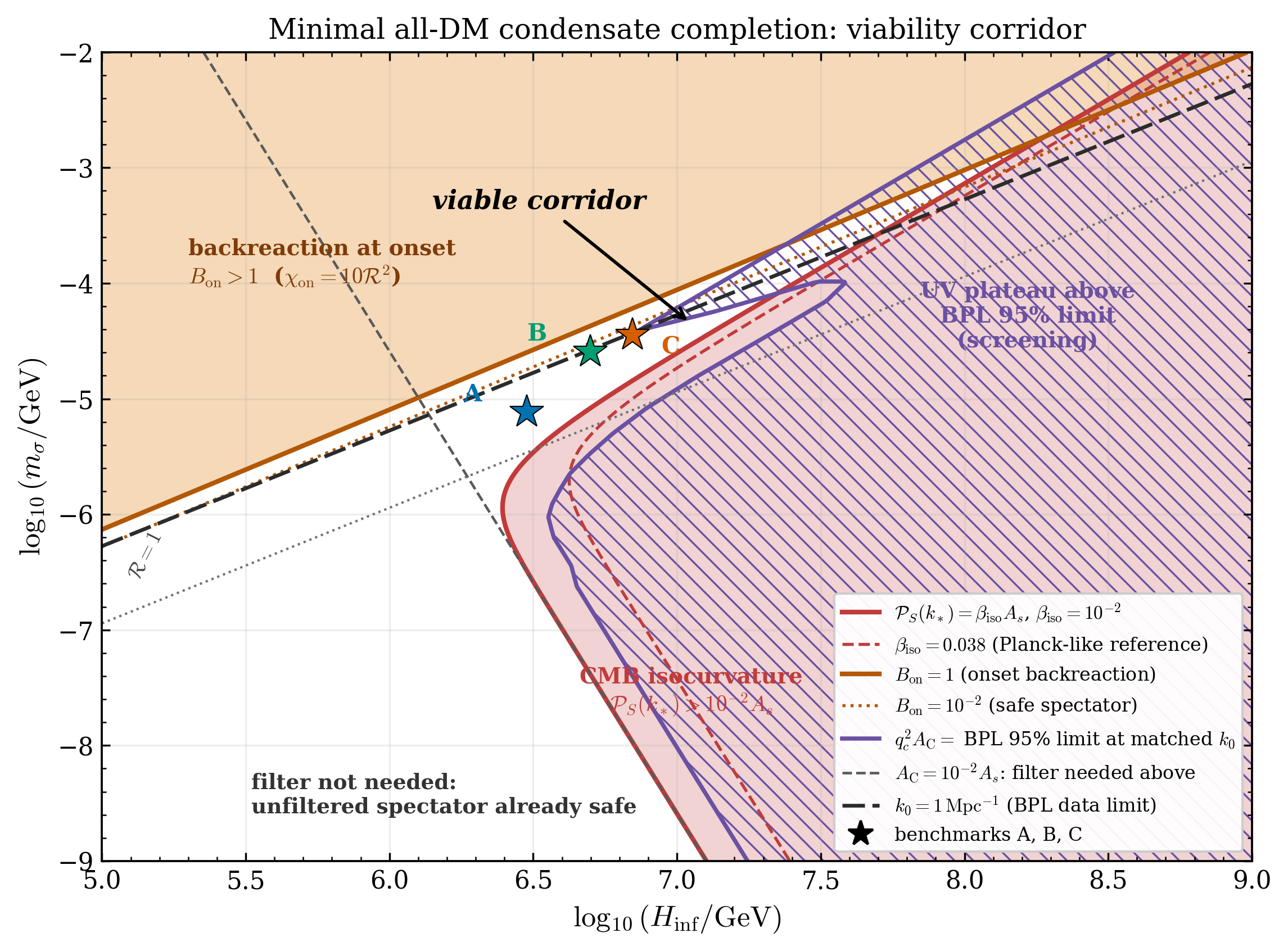}
 \caption{\small
  Viability corridor of the minimal all-dark-matter condensate completion
  ($f_\sigma=1$) in the $(\hinf,\ms)$ plane.  \emph{Red}: excluded by the exact
  pivot condition $\Ps(\ks)>10^{-2}A_s$ (solid; dashed shows the Planck-like
  reference $\beta_{\rm iso}=0.038$).  At fixed $\hinf$ this is a \emph{band},
  because $\Ps(\ks)$ is non-monotonic in $\ms$,
  Eq.~\eqref{eq:corridor-pivot-scalings}.  \emph{Orange}: homogeneous
  backreaction at the onset of the filter, $B_{\rm on}>1$, for the reference
  prescription $\chi_{\rm on}=10 R^{2}$ and $c_{\rm bg}=1.3$ (dotted:
  $B_{\rm on}=10^{-2}$).  \emph{Hatched}: ultraviolet plateau above the digitized
  $95\%$ broken-power-law limit at the matched break scale $k_0=0.923\,\kc$;
  by Eq.~\eqref{eq:pointwise-bpl-bound} this screening is pointwise conservative
  for the smooth Coulomb shape.  \emph{Gray dashed}: $A_{\rm C}=10^{-2}A_s$; below
  this line the unfiltered spectator is already safe and the filter is not needed,
  so only the region above it is claimed as a filtering solution.  Thin dotted
  lines mark $R =1$ and $\kc=10\,\mathrm{Mpc}^{-1}$.  The corridor spans
  $10^{6}\lesssim\hinf/\mathrm{GeV}\lesssim2\times10^{7}$; at the upper end the
  binding constraint is the plateau screening rather than backreaction (see text).
  Stars: the benchmarks of Table~\ref{tab:coulomb-benchmarks}.
 }
 \label{fig:benchmark-exact-spectra}
\end{figure}

\paragraph{The $(\hinf,\ms)$ plane.}
The three benchmarks of Table~\ref{tab:coulomb-benchmarks} are individual points
in a two-parameter family.  Once the inflationary inputs $N_*$, $\ks$ and $A_s$
are fixed, and once $\sigma$ is required to be \emph{all} of the dark matter
($f_\sigma=1$, so that $\sigbar_{\rm DM}$ is determined by $\ms$ through the
abundance relation~\eqref{eq:sigma_DM}), every quantity entering our constraints
is a function of $\hinf$ and $\ms$ alone.  It is therefore useful to display the
constraints directly in that plane, which is what Fig.~\ref{fig:benchmark-exact-spectra} does.
Three derived quantities control the whole picture,
\begin{equation}
  R=e^{N_*}\Big(\frac{\ms}{\hinf}\Big)^{2},
  \qquad
  A_{\rm C}=\Big(\frac{\hinf}{\pi\sigbar_{\rm DM}}\Big)^{2}\propto\hinf^{2}\ms^{1/2},
  \qquad
  \Ps(\ks)=A_{\rm C}\,\PT(\ks;\kc) ,
  \label{eq:corridor-three-quantities}
\end{equation}
where the $\ms^{1/2}$ in $A_{\rm C}$ comes from the misalignment normalization
$\sigbar_{\rm DM}\propto\ms^{-1/4}$, see Eq.~\eqref{eq:misalignment_abundance}. A lighter condensate needs a \emph{larger}
displacement to carry the observed abundance, and therefore has \emph{smaller}
fractional isocurvature.  This  fact is responsible for most of the
non-trivial shape of the figure.

\paragraph{\it Constructing  the shaded regions:}
Four conditions are evaluated on a grid in $(\log_{10}\hinf,\log_{10}\ms)$ and
displayed in Fig.~\ref{fig:benchmark-exact-spectra}.  Three of them are exclusions,
\begin{eqnarray}
&&  \text{(i) large-scale isocurvature:}\quad
 \Ps(\ks)\;>\;\beta_{\rm iso}A_s,
  \qquad \beta_{\rm iso}=10^{-2},
  \label{eq:corridor-cmb}\\[2pt]
 && \text{(ii) homogeneous backreaction:}\quad
  B_{\rm on}\;>\;1,
  \qquad \chi_{\rm on}=10\,R^{2},\ c_{\rm bg}=1.3,
  \label{eq:corridor-back}\\[2pt]
  && \text{(iii) plateau screening:}\quad
  q_c^{2}A_{\rm C}\;>\;q_c^{2}A_{\rm iso}^{95}(k_0),
  \qquad k_0=(\pi/4)^{1/3}\kc,
  \label{eq:corridor-bpl}
\end{eqnarray}
shaded in red, orange and hatched purple respectively, while the fourth,
\begin{eqnarray}
 \hskip-3cm \text{(iv) relevance of the filter:}\qquad
  A_{\rm C}\;>\;10^{-2}A_s ,
  \label{eq:corridor-relevance}
\end{eqnarray}
is \emph{not} an exclusion: it is the requirement (discussed above)  that the unfiltered spectator
spectrum would itself violate the design level, so that the Coulomb filter is
doing genuine work.  It is drawn as a gray dashed line, and only the region above
it is claimed as a filtering solution.  In~\eqref{eq:corridor-cmb} the pivot power
is computed with the \emph{exact} transfer function~\eqref{eq:coulomb-transfer}
rather than its infrared limit, which matters near the boundary because
$\PT(\kc)\simeq0.24$ rather than $4/\pi$.  In~\eqref{eq:corridor-bpl},
$A_{\rm iso}^{95}(k_0)$ is the digitized $95\%$ upper limit of
Ref.~\cite{Buckley:2025zgh} on the broken-power-law amplitude, evaluated at the
break scale matched to our spectrum through
Eq.~\eqref{eq:benchmark-asymptotic-map}.  The union of
\eqref{eq:corridor-cmb}--\eqref{eq:corridor-bpl}, intersected with
\eqref{eq:corridor-relevance}, leaves the white wedge labelled \emph{viable
corridor}; a point may of course be excluded by more than one condition, and the
shadings are allowed to overlap.

\paragraph{\it Why the isocurvature exclusion is a band and not a half-plane?}
The most counter-intuitive feature of Fig.~\ref{fig:benchmark-exact-spectra} is that at fixed
$\hinf$ the red region is bounded both above and below in $\ms$.  The reason is
that $\Ps(\ks)$ is not monotonic in the mass.  Using
Eq.~\eqref{eq:corridor-three-quantities} and the asymptotics
\eqref{eq:benchmark-coulomb-asymptotics},
\begin{equation}
  \Ps(\ks)\;\propto\;
  \begin{cases}
    \hinf^{2}\,\ms^{1/2}, & R\lesssim1 \quad\text{(pivot on the plateau)},\\[6pt]
    \hinf^{8}\,\ms^{-11/2}, & R\gg1 \quad\text{(pivot on the $k^{3}$ branch)} .
  \end{cases}
  \label{eq:corridor-pivot-scalings}
\end{equation}
Increasing $\ms$ at fixed $\hinf$ therefore has two competing effects. It raises
the plateau amplitude as $\ms^{1/2}$, but it also pushes the break scale
$\kc=R\ks\propto\ms^{2}$ to the ultraviolet, and once the pivot sits on the
$k^{3}$ branch the suppression $R^{-3}\propto\ms^{-6}$ wins by a wide
margin.  The pivot power consequently peaks at an intermediate mass and falls off
steeply on either side.  Numerically the maximum occurs at $R\simeq0.16$,
i.e.\ for a break scale slightly \emph{below} the pivot,
$\kc\simeq8\times10^{-3}\,\mathrm{Mpc}^{-1}$ --- the exact transfer function
already departs from unity for $u=\kc/k\sim0.1$, so the turnover of $\Ps(\ks)$
sets in before the break formally crosses $\ks$.  Reading the figure vertically at
fixed $\hinf$ one therefore passes through four distinct regimes: an unfiltered
but harmless region at small $\ms$ (below the gray dashed line, where the
condensate is light, $\sigbar_{\rm DM}$ is large and $A_{\rm C}$ is small); the
excluded band, where the break has not yet moved far enough into the ultraviolet
to protect the pivot; the corridor, where the filter is both needed and
sufficient; and finally the backreaction region.  Only the third of these is a
filtering solution in the sense of our mechanism.

\paragraph{\it Upper boundary: backreaction.}
The corridor is closed from above by the onset backreaction
condition~\eqref{eq:corridor-back}.  With the reference prescription
$\chi_{\rm on}=10 R^{2}$ one has, from Eq.~\eqref{eq:benchmark-Bon},
\begin{equation}
  B_{\rm on}\;\propto\;\sigbar_{\rm DM}^{2}\,\chi_{\rm on}^{7/2}
  \;\propto\;\ms^{27/2}\,\hinf^{-14} ,
  \label{eq:corridor-back-scaling}
\end{equation}
a very steep dependence on the mass.  Two consequences:
First, the boundary is a nearly straight line of slope
$\mathrm{d}\log\ms/\mathrm{d}\log\hinf=28/27\simeq1.04$ in the log-log plane.
Second, and more importantly, its \emph{location} is almost independent of the
threshold one chooses: tightening the requirement from $B_{\rm on}<1$ to
$B_{\rm on}<10^{-2}$ moves the boundary down in mass by only
$10^{2/13.5}\simeq1.4$.    The steepness cuts both ways: the same
exponent means that $B_{\rm on}$ itself is only meaningful as an order-of-magnitude
diagnostic, since it inherits the full $\chi_{\rm on}^{7/2}$ sensitivity to the
onset prescription.  What is robust is the boundary in $\ms$, not the value of
$B_{\rm on}$ at a given point.

\paragraph{\it Extent of the corridor and the constraint that closes it.}
The corridor is bounded in $\hinf$ at both ends.  At low $\hinf$ it does not exist
at all: the relevance condition~\eqref{eq:corridor-relevance} requires
$\ms\gtrsim\hinf^{-4}$ (the gray dashed line falls steeply to the right, since
$A_{\rm C}\propto\hinf^{2}\ms^{1/2}$), whereas backreaction requires
$\ms\lesssim\hinf^{28/27}$, and the two are incompatible below
\begin{equation}
  \hinf\simeq1\times10^{6}\ \mathrm{GeV} .
  \label{eq:corridor-opening}
\end{equation}
Below this scale a spectator that carries all the dark matter is simply safe
without any filtering, and the mechanism is not needed.  At the opposite end the
corridor closes at
\begin{equation}
  \hinf\simeq2\times10^{7}\ \mathrm{GeV},
  \label{eq:corridor-closure}
\end{equation}
with $\ms\sim10^{-4}$~GeV near the endpoint.  Geometrically the closure happens
because the lower boundary rises faster than the upper one: on the filtered branch
the isocurvature boundary has slope
$\mathrm{d}\log\ms/\mathrm{d}\log\hinf=16/11\simeq1.45$ from
Eq.~\eqref{eq:corridor-pivot-scalings}, against $28/27\simeq1.04$ for
backreaction, so the wedge pinches shut.  Physically, raising $\hinf$ raises the
unfiltered amplitude as $\hinf^{2}$ and must be compensated by a larger $R$,
hence a larger mass. But a larger mass costs backreaction at the seventh power of
$\chi_{\rm on}$, and simultaneously pushes the plateau up as $\ms^{1/2}$ into the
reach of the small-scale limits.  Which of the two upper constraints actually
binds at the tip depends mildly on how conservative one is. Imposing only
$B_{\rm on}<1$, the endpoint is set by backreaction, with the plateau screening
within $\sim10\%$ of saturation; imposing the much stronger $B_{\rm on}<10^{-2}$,
the endpoint moves only to $\hinf\simeq2.1\times10^{7}$~GeV and is then set by the
plateau screening alone.  In other words, the closure is controlled by the
small-scale power of the ultraviolet plateau --- an observational statement,
already probed by existing broken-power-law likelihoods --- and not by spectator
self-consistency, which is the more model-dependent of the two.  For orientation,
at $\hinf=10^{7}$~GeV the corridor spans
$\ms\simeq2.5\times10^{-5}$--$8.8\times10^{-5}$~GeV, a factor of about three in
mass. The three benchmarks of Table~\ref{tab:coulomb-benchmarks} sit comfortably
inside it. To conclude, after imposing all our requirements, the inflationary
scale turns out to be relatively low, and the final spectator mass around the keV range. Going beyond the minimal scenario discussed here we can modify these
results, at the price of losing in predictivity. We will discuss possible
options in the Outlook section \ref{sec:outlook}.

\subsection{Relation to other blue-isocurvature mechanisms}
\label{sec:context}

The Coulomb filter discussed in this work belongs to a broader class of mechanisms in which the
dark-matter isocurvature spectrum is made strongly blue by a time-dependent
spectator sector \cite{Kasuya:2009up,Chung:2015tha,Chung:2015pga,Chung:2016wvv,Chung:2017uzc,Chung:2021lfg,Chung:2024ctx}.  It is useful to state precisely what is common to these
models and what is specific to our construction.

\paragraph{A large blue index as evidence for time-dependent spectator
dynamics.}
We define the isocurvature spectral index by
\begin{equation}
 {\cal P}_{S}(k)\propto k^{n_I-1}
\end{equation}
within a range of scales over which the spectrum is approximately a power
law.  For a linear spectator field with a constant mass throughout the
relevant inflationary and post-inflationary history, Ref.~\cite{Chung:2015tha,Chung:2015pga}
shows that the largest \emph{measurable} blue index compatible with the
tensor bound is approximately
\begin{equation}
 n_I\simeq 2.4.
 \label{eq:context-constant-mass-bound}
\end{equation}
The numerical value is not a universal theorem about every possible
isocurvature model.  It assumes a linear spectator, a constant mass over the
relevant history, an observable but subdominant isocurvature component, and
the tensor constraint used in that analysis.  Its robust qualitative lesson
is nevertheless important: an observed spectrum appreciably bluer than this
range would point towards an evolving mass, an evolving normalization of the
isocurvature degree of freedom, or a non-trivial conversion mechanism~\footnote{
General results concerning the conservation and quantization of linear
spectator isocurvature perturbations were developed in
Ref.~\cite{Chung:2015pga}.}
 The Coulomb filter is therefore an explicit and analytically soluble
realization of the general expectation that a very blue linear isocurvature
spectrum requires time-dependent spectator dynamics.

There is also an important distinction between a pure power law and the
physical spectrum obtained in our work.  The \(k^3\) behaviour persists only below
the turnover.  At \(k\gg \kc\), the transfer function tends to a constant.
A blue branch connected to a short-scale plateau is common in models where a
time-dependent mass or field normalization eventually settles
\cite{Chung:2016wvv,Chung:2017uzc}.  The precise shape of the transition is
model-dependent.  The distinctive feature of the present construction is
that both the envelope and the transition are known in closed and elegant
form, given by Eq.~\eqref{eq:coulomb-transfer}.

\paragraph{Relation to underdamped and resonant blue isocurvature.}
The physical mass during the Coulomb stage satisfies
\begin{equation}
 \frac{M_\sigma^2}{H_{\rm inf}^2}
 =
 \kc(-\tau)\equiv\chi(\tau).
 \label{eq:context-chi}
\end{equation}
At horizon crossing of a mode \(k\),
\begin{equation}
 \chi_k\simeq\frac{\kc}{k}.
 \label{eq:context-chi-crossing}
\end{equation}
Modes well inside the filtered region, \(k\ll \kc\), consequently encounter
a heavy and underdamped spectator sector.  More precisely, the homogeneous
motion is underdamped when \(M_\sigma>3H_{\rm inf}/2\), which is amply
satisfied for \(\chi\gg1\).

The underdamped regime of blue axion isocurvature was studied analytically in
Ref.~\cite{Chung:2021lfg}, finding that oscillatory background
dynamics and time-dependent mixing among several axionic degrees of freedom
can generate resonant enhancements, dips, and a rich transition region. 
This provides a useful comparison for the finite-duration versions of the
present model.  A finite pulse tower, an abrupt onset, or a rapid termination of
the Coulomb stage introduces additional temporal scattering and produces
oscillatory ringing around the smooth envelope; such
ringing can accompanied by a genuine distortion of the infrared slope rather than
by a modulation about it.  This is one reason we require the onset to be
adiabatic, Eq.~\eqref{eq:onset-adiabatic-condition}: the smooth Coulomb result
is not merely a convenient idealization of a sharp switch-on, but a distinct
physical regime.

The two mechanisms should not, however, be identified.  The resonances of
Ref.~\cite{Chung:2021lfg} arise from non-perturbative classical evolution and
mixing in a multi-field Peccei--Quinn sector.  The universal Coulomb spectrum
instead follows from a single linear mode equation with the smooth profile
\(a^2M_\sigma^2=\kc/(-\tau)\).  The ideal infinite-duration profile does not
require a background resonance.  Resonant or oscillatory features appear
only when the profile is completed by finite endpoints, discrete pulses, or
additional fields.  The underdamped axion analysis is therefore best viewed
as a source of analytic techniques and physical intuition for such
extensions, rather than as a derivation of the Coulomb result itself.

\paragraph{Comparison with geometric suppression.}
A  close recent comparison is the hyperbolic axion construction
of Ref.~\cite{Tadepalli:2026mdc}.  In that model, the curved metric of the
Peccei--Quinn field space gives the canonically normalized angular
fluctuation a time-dependent geometric mass.  The same geometry enhances the
effective angular normalization, suppressing CMB-scale isocurvature while
producing a blue spectrum on shorter scales.  The proposal does not require
explicit Peccei--Quinn breaking or an extreme radial displacement and
exhibits benchmarks with
\begin{equation}
 H_{\rm inf}\sim10^{13}\ {\rm GeV}.
\end{equation}

Both constructions use an evolving
inflationary spectator sector to suppress long-wavelength isocurvature, and
both predict a blue spectrum whose scale dependence contains information
about that evolution.  Their theoretical organizations are nevertheless
different.  In the hyperbolic model, the time dependence follows from a
specified field-space geometry and from the radial background trajectory.
In the present model, the primary object is instead the exactly soluble
Coulomb profile, which may be understood either as the continuum limit of a
logarithmic pulse tower or as the effective mass generated by an auxiliary
background.

A complementary strategy is to give the spectator a large mass during
inflation which is subsequently removed. Ref.~\cite{Chakraborty:2025lyp} realizes
this through a type of monodromy mass whose integer coefficient is discharged
by brane nucleation, so that the mass is large during inflation and exactly
zero afterwards. The mechanism developed here differs in that the mass is not
switched between two values but graded continuously as $a^{-1/2}$ across the
filtering stage; it is this grading, rather than the overall size of the mass,
that produces our spectrum with a single smooth turnover instead of an abrupt
suppression scale.


\section{Outlook}
\label{sec:outlook}

The main result of this work is more general than the particular dark-matter
realization used to illustrate it.  A logarithmically ordered sequence of
localized mass perturbations admits a dense limit in which the spectator mode
equation becomes an exactly soluble Coulomb problem.  The resulting transfer
function,
\begin{equation}
 {\cal T}_{\rm C}(k)=
 \frac{1-e^{-\pi\kappa_c/k}}
 {(\pi\kappa_c/k)\left[1+\kappa_c^2/(4k^2)\right]},
 \label{eq:outlook-transfer}
\end{equation}
provides a smooth interpolation between a strongly suppressed infrared branch,
${\cal P}_{\delta\sigma}\propto k^3$, and the standard massless-spectator
plateau.  This result depends neither on a dark-matter abundance prescription
nor on the subsequent cosmological history of the spectator.  In this sense,
the Coulomb filter should be regarded as the universal part of the
construction, while the dark-matter scenario of
Sec.~\ref{sec:filtered-spectator-dm} is one particularly economical completion.

Several features make the filtered spectrum distinctive.  First, its shape is
essentially controlled by a single scale, $\kappa_c$: once the overall
normalization is fixed, the infrared coefficient, the turnover region, and the
ultraviolet plateau are not independent.  The smooth turnover therefore obeys
a non-trivial shape consistency relation that distinguishes the Coulomb
profile from a generic broken-power-law parametrization.  Second, the
continuum limit contains no intermediate enhancement or bump.  The dips and
oscillatory structures produced by a finite number of temporal scatterers
average away, leaving a monotonic high-pass filter.  A detected spectrum with a
smooth $k^3$ rise and no accompanying overshoot would consequently point
toward the coarse-grained regime of the construction rather than toward a
small number of isolated events.

Conversely, departures from the ideal limit can carry information about the
microscopic origin of the filter.  A finite number of pulses, a finite onset or
termination time, or a clock frequency that is large but not infinitely
separated from the frequencies probed by the spectator leave residual
oscillations around the Coulomb envelope.  For the clock realization discussed
above, these corrections are approximately periodic in $\ln k$.  Such
log-periodic ringing is therefore one of the most pulse-specific signatures of
the mechanism: the smooth Coulomb profile captures the universal envelope,
whereas the residual oscillations can retain information about the temporal
spacing and duration of the underlying events.  It would be interesting to
develop the finite-window problem analytically and determine how much of this
microscopic information can survive in observable isocurvature spectra.

A further attractive feature is that the filtering takes place directly in the
spectator sector.  In the simplest realization the inflaton dynamics, and hence
the primordial adiabatic spectrum, need not contain a corresponding feature.
The mechanism can therefore generate a pronounced scale dependence in
isocurvature while leaving the curvature perturbation close to the usual
slow-roll form.  This offers a useful phenomenological discriminator from
scenarios in which the same inflationary feature simultaneously imprints both
adiabatic and isocurvature perturbations.

The minimal dark-matter completion developed in this paper adds another
potentially observable relation.  Under the assumptions adopted in
Sec.~\ref{sec:filtered-spectator-dm}, continuity of the spectator mass profile
at the end of inflation gives
\begin{equation}
 \frac{\kappa_c}{k_*}
 =
 e^{N_*}
 \left(\frac{m_\sigma}{H_{\rm inf}}\right)^2 .
 \label{eq:outlook-mass-turnover}
\end{equation}
The turnover scale is therefore not an arbitrary phenomenological parameter:
it is directly tied to the late-time dark-matter mass.  Together with the
misalignment abundance relation, Eq.~\eqref{eq:outlook-mass-turnover} makes the
minimal realization highly predictive.  If evidence for a blue isocurvature
component with a turnover were ever established, this relation would provide a
direct consistency test between its spectral location and the microscopic
parameters of the spectator sector.

It is important, however, to distinguish this predictive minimal realization
from the filtering mechanism itself.  The numerical viability corridor found
in the $(H_{\rm inf},m_\sigma)$ plane follows from a deliberately restrictive
set of assumptions: the homogeneous displacement is already present while the
filter operates, the spectator constitutes all of the dark matter, the
Coulomb-like mass evolution is matched directly onto a constant post-inflationary
mass, and the subsequent abundance is generated by the standard quadratic
misalignment mechanism.  These assumptions make the system economical and
allow essentially every relevant constraint to be followed analytically, but
they should not be interpreted as necessary ingredients of Coulomb-filtered
dark matter.  In particular, the upper range of inflationary scales obtained in
our benchmarks is not a model-independent upper bound on the mechanism.

This distinction becomes especially important for the homogeneous
backreaction constraint.  In the minimal setup, the condensate is already
present during the early heavy-mass stage and its energy density grows rapidly
toward the onset of the filter when evolved backwards in time.  The resulting
constraint is therefore sensitive to the assumed initial displacement and to
the precise time at which the Coulomb stage begins.  More elaborate
post-inflationary dynamics can change this conclusion substantially.  For
example, the large homogeneous displacement responsible for the present dark
matter abundance could be generated only near the end of inflation or during
reheating, after the filtering of the primordial fluctuations has already
occurred.  A reheating-induced displacement of the minimum, a transient
tadpole, mixing with another scalar condensate, or a change of the effective
spectator potential could all generate or amplify the homogeneous mode without
repopulating the long-wavelength fluctuations that were filtered during
inflation.  In such cases the dominant heavy-condensate backreaction constraint
of the minimal realization can be strongly relaxed or absent altogether.

More generally, reheating introduces several physically independent effects.
A noninstantaneous reheating stage changes the relation between $N_*$ and the
post-inflationary thermal history, modifies the mapping between $\kappa_c$ and
physical scales today, and can alter both the spectator mass and the onset of
its late-time oscillations.  If the spectator couples, even weakly, to the
reheating sector, its effective potential can evolve non-adiabatically across
this epoch.  The final condensate amplitude then need not be identical to the
value frozen at the end of inflation, and the simple mass--turnover relation
\eqref{eq:outlook-mass-turnover} can be replaced by a more general transfer
relation containing information about reheating.  Rather than being a
drawback, this opens the possibility that an observed isocurvature turnover
could probe not only the spectator mass but also the otherwise poorly known
intermediate expansion history.

There are several natural directions in which the dark-matter completion can
therefore be generalized:
\begin{itemize}[leftmargin=1.6em]

 \item \emph{Dynamical condensate generation.}
 The homogeneous mode may be produced only at the end of inflation or during
 reheating, rather than pre-existing throughout the Coulomb stage.  This
 separates the mechanism responsible for filtering the fluctuations from the
 mechanism responsible for generating the dark-matter abundance.

 \item \emph{Non-quadratic spectator dynamics.}
 Self-interactions, a moving minimum, anharmonic evolution, or a transition
 between different effective potentials can modify the relation between the
 frozen field amplitude and the late-time abundance.  The spectral filter can
 remain linear even when the homogeneous condensate subsequently evolves
 non-linearly.

 \item \emph{A non-standard reheating history.}
 An extended matter-like, kination-like, or otherwise non-radiation-dominated
 phase changes the abundance relation, the value of $N_*$ associated with a
 given observed scale, and the mapping between the inflationary turnover and
 the final dark-matter mass.

 \item \emph{A subdominant spectator component.}
 If $\sigma$ constitutes only a fraction of the dark matter, both the observable
 isocurvature amplitude and the homogeneous backreaction are weakened.  The
 same filtering mechanism can therefore operate in a considerably wider region
 of parameter space.

 \item \emph{Finite-duration and finite-density filters.}
 A smooth turn-on and turn-off, or a pulse spacing that is small but finite,
 replaces the ideal Coulomb result by a calculable deformation containing
 endpoint and log-periodic information.  These corrections may provide a way
 of reconstructing properties of the underlying clock or pulse-generating
 sector.
\item \emph{Extensions to other spins,} as for example primordial
gravitational waves \cite{Tasinato:2026fgu}. 
\end{itemize}

These extensions emphasize a useful separation of roles.  The inflationary
filter determines the \emph{shape} and large-scale suppression of the
primordial spectator fluctuations; the later condensate dynamics determine
how those fluctuations are converted into a dark-matter abundance and into the
observable isocurvature mode.  In the minimal model these two stages are tied
together as tightly as possible, which is why the turnover, abundance, mass,
and backreaction constraints become strongly correlated.  In a more general
completion they need not be.  The filtered spectrum of
Eq.~\eqref{eq:outlook-transfer} can therefore survive even when the numerical
dark-matter bounds derived in Sec.~\ref{sec:filtered-spectator-dm} are
substantially modified.

From this perspective, the broader lesson of our construction is that a very
blue primordial spectrum need not arise from a resonance or from intrinsically
non-linear dynamics.  It can instead emerge from a simple linear
temporal-scattering problem whose self-similar dense limit is exactly soluble.
The logarithmic pulse tower provides a microscopic interpretation of the
Coulomb profile, while the Coulomb solution exposes the universal behaviour
hidden in the many-event system.  The combination of a causal $k^3$ infrared
branch, a smooth one-parameter turnover, possible log-periodic residuals, and,
in minimal dark-matter completions, a direct mass--turnover relation provides a
set of signatures that can be searched for independently of the details of the
ultraviolet realization.

It would be particularly interesting to confront the exact Coulomb shape
directly with cosmological data rather than through broken-power-law
proxies, and to determine to what extent future measurements of small-scale
isocurvature can distinguish the universal smooth envelope from finite-pulse
residuals.  More broadly, the temporal-scattering viewpoint developed here can
be applied to other spectator sectors and to other inflationary observables.
The present construction provides one example in which an apparently
complicated sequence of localized events reorganizes itself into a simple,
predictive, and exactly soluble primordial filter.

\subsection*{Acknowledgments}
It is a pleasure to thank Susha Parameswaran and Ivonne Zavala for 
useful input. GT is partially funded by the STFC grant ST/X000648/1, by the Royal Society grant IES  R3  43186 and by the Leverhulme Trust grant RF-2026-166 9.
MLR acknowledges support by the Italian MUR through the FIS 2 project FIS-2023-01577 (DD n. 23314 10-12-2024, CUP C53C24001460001), and by Istituto Nazionale di Fisica Nucleare (INFN) through the Theoretical Astroparticle Physics (TAsP) project.
\appendix

\section{Finite-pulse transfer matrices}
\label{app:one-two-kicks}

This appendix collects the exact matching formulae for an arbitrary finite
sequence of instantaneous spectator-mass kicks.  It also gives the one- and
two-kick results used in the main text.

Between successive kicks, the canonically normalized spectator mode obeys
the massless de Sitter equation.  We use the global basis
\begin{equation}
 f_k(\tau)
 =
 \frac{1}{\sqrt{2k}}
 \left(1-\frac{\ii}{k\tau}\right)e^{-\ii k\tau},
 \qquad
 f_k f_k^{*\prime}-f_k'f_k^*=\ii .
 \label{eq:app-BD-basis}
\end{equation}
Consider a positive pulse of area \(s_j\) at
\(\tau_j=-T_j\), with \(T_j>0\):
\begin{equation}
 a^2\delta M_{\sigma,j}^2(\tau)
 =
 s_j\,\delta(\tau+T_j).
 \label{eq:app-pulse-profile}
\end{equation}
The mode and its derivative satisfy
\begin{equation}
 [u_k]_j=0,
 \qquad
 [u_k']_j=-s_j u_k(\tau_j).
 \label{eq:app-jump}
\end{equation}

Immediately before and after the kick, write
\begin{equation}
 u_k^-
 =
 A_j^- f_k+B_j^- f_k^*,
 \qquad
 u_k^+
 =
 A_j^+ f_k+B_j^+ f_k^*.
 \label{eq:app-mode-decomposition}
\end{equation}
Using the Wronskian in Eq.~\eqref{eq:app-BD-basis}, the jump condition gives
\begin{equation}
 \begin{pmatrix}
 A_j^+\\[2pt] B_j^+
 \end{pmatrix}
 =
 \mathsf K_j
 \begin{pmatrix}
 A_j^-\\[2pt] B_j^-
 \end{pmatrix},
 \label{eq:app-kick-map}
\end{equation}
where
\begin{equation}
 \mathsf K_j
 =
 \begin{pmatrix}
 1-\ii s_j|f_j|^2 & -\ii s_j f_j^{*\,2}\\[2pt]
 \ii s_j f_j^2 & 1+\ii s_j|f_j|^2
 \end{pmatrix}
 =
 \begin{pmatrix}
 a_j & b_j^*\\
 b_j & a_j^*
 \end{pmatrix},
 \qquad
 f_j\equiv f_k(-T_j).
 \label{eq:app-kick-matrix-general}
\end{equation}
Defining
\begin{equation}
 x_j\equiv kT_j,
 \qquad
 q_j\equiv s_jT_j,
 \label{eq:app-xq}
\end{equation}
the matrix elements become
\begin{align}
 a_j
 &=
 1-\frac{\ii q_j}{2x_j}
 \left(1+\frac{1}{x_j^2}\right),
 \label{eq:app-aj}
 \\
 b_j
 &=
 \frac{\ii q_j}{2x_j}
 \left(1+\frac{\ii}{x_j}\right)^2e^{2\ii x_j}.
 \label{eq:app-bj}
\end{align}
For a positive mass kick, \(s_j>0\) and hence \(q_j>0\).

Each kick preserves the canonical Wronskian:
\begin{equation}
 |a_j|^2-|b_j|^2=1,
 \qquad
 \det\mathsf K_j=1.
 \label{eq:app-SU11}
\end{equation}

\paragraph{One kick}
\label{app:one-kick}

For an incoming Bunch--Davies mode,
\begin{equation}
 \begin{pmatrix}\alpha_1\\[2pt]\beta_1\end{pmatrix}
 =
 \mathsf K_1
 \begin{pmatrix}1\\[2pt]0\end{pmatrix}
 =
 \begin{pmatrix}a_1\\[2pt]b_1\end{pmatrix}.
 \label{eq:app-one-vector}
\end{equation}
As \(\tau\to0^-\), the constant mode of \(\delta\sigma=u/a\) is proportional
to \(\alpha_1-\beta_1\).  The exact one-kick transfer function is therefore
\begin{equation}
 {\cal T}_1(x;q)
 =
 |\alpha_1-\beta_1|^2.
 \label{eq:app-one-transfer-definition}
\end{equation}
Introducing
\begin{equation}
 c_0(x)
 =
 1+x^2+e^{2\ii x}(x+\ii)^2,
 \label{eq:app-c0}
\end{equation}
one obtains
\begin{equation}
 \alpha_1-\beta_1
 =
 1-\frac{\ii q}{2x^3}c_0(x),
\end{equation}
and hence
\begin{equation}
 {
 {\cal T}_1(x;q)
 =
 \left|
 1-\frac{\ii q}{2x^3}c_0(x)
 \right|^2 .
 }
 \label{eq:app-one-exact}
\end{equation}

\paragraph{Two kicks and interference}
\label{app:two-kicks}

Let the first pulse occur at \(-T_1\) and the second at \(-T_2\), with
\begin{equation}
 T_1>T_2>0.
\end{equation}
The exact final coefficients are
\begin{equation}
 \begin{pmatrix}\alpha_{12}\\[2pt]\beta_{12}\end{pmatrix}
 =
 \mathsf K_2\mathsf K_1
 \begin{pmatrix}1\\[2pt]0\end{pmatrix},
 \label{eq:app-two-vector}
\end{equation}
or
\begin{align}
 \alpha_{12}
 &=
 a_2a_1+b_2^*b_1,
 \label{eq:app-two-alpha}
 \\
 \beta_{12}
 &=
 b_2a_1+a_2^*b_1.
 \label{eq:app-two-beta}
\end{align}
The late-time transfer function is
\begin{equation}
 {
 {\cal T}_2(k)
 =
 \left|
 a_2a_1+b_2^*b_1-b_2a_1-a_2^*b_1
 \right|^2 .
 }
 \label{eq:app-two-transfer}
\end{equation}
This expression includes both direct interference and rescattering of the
negative-frequency component produced by the first kick.

For weak pulses acting while the mode is well inside the horizon,
\begin{equation}
 x_j\gg1,
 \qquad
 \frac{s_j}{k}\ll1,
\end{equation}
the matrix elements reduce to
\begin{align}
 a_j
 &=
 1-\frac{\ii s_j}{2k}
 +{\cal O}\!\left(\frac{s_j}{kx_j^2}\right),
 \\
 b_j
 &=
 \frac{\ii s_j}{2k}e^{2\ii x_j}
 +{\cal O}\!\left(\frac{s_j}{kx_j}\right).
\end{align}
To first order in the pulse areas,
\begin{equation}
 \beta_{12}(k)
 \simeq
 \frac{\ii}{2k}
 \left[
 s_1e^{2\ii kT_1}
 +
 s_2e^{2\ii kT_2}
 \right],
 \label{eq:app-two-beta-weak}
\end{equation}
and therefore
\begin{equation}
 |\beta_{12}|^2
 \simeq
 \frac{
 s_1^2+s_2^2+
 2s_1s_2\cos\!\left[2k(T_1-T_2)\right]
 }{4k^2}.
 \label{eq:app-two-interference}
\end{equation}
For equal pulse areas,
\begin{equation}
 |\beta_{12}|^2
 \simeq
 \frac{s^2}{k^2}
 \cos^2\!\left[k(T_1-T_2)\right].
\end{equation}
The pulse separation fixes the oscillation period, while the relative areas
set the interference contrast.

\paragraph{An arbitrary finite tower}
\label{app:N-kicks}

For \(N\) pulses ordered as
\begin{equation}
 T_1>T_2>\cdots>T_N>0,
\end{equation}
the exact transfer matrix is
\begin{equation}
 \mathsf K_{\rm tot}
 =
 \mathsf K_N\mathsf K_{N-1}\cdots\mathsf K_1
 =
 \begin{pmatrix}
 A_N & B_N^*\\
 B_N & A_N^*
 \end{pmatrix}.
 \label{eq:app-total-matrix}
\end{equation}
For a Bunch--Davies incoming state,
\begin{equation}
 \begin{pmatrix}\alpha_N\\[2pt]\beta_N\end{pmatrix}
 =
 \mathsf K_{\rm tot}
 \begin{pmatrix}1\\[2pt]0\end{pmatrix}
 =
 \begin{pmatrix}A_N\\[2pt]B_N\end{pmatrix},
\end{equation}
and
\begin{equation}
 {
 {\cal T}_N(k)
 =
 |A_N-B_N|^2.
 }
 \label{eq:app-N-transfer}
\end{equation}
Because the basis \(f_k,f_k^*\) is defined globally between all pulses, the
free propagation phases are already contained in the factors
\(e^{2\ii kT_j}\) appearing in \(b_j\).  No separate propagation matrix is
required in this convention.

For the logarithmic tower used in the main text,
\begin{equation}
 T_j=T_{\rm on}e^{-(j-1)h},
 \qquad
 s_j=h \kc.
 \label{eq:app-log-tower}
\end{equation}
The pulse area is constant, whereas the dimensionless strength
\begin{equation}
 q_j=s_jT_j=h \kc T_j
\end{equation}
decreases towards the end of inflation.  At fixed physical interval
\([T_{\rm off},T_{\rm on}]\), the number of pulses scales as
\begin{equation}
 N\simeq
 \frac{1}{h}
 \ln\!\left(\frac{T_{\rm on}}{T_{\rm off}}\right).
\end{equation}
As \(h\to0\), the ordered product in Eq.~\eqref{eq:app-total-matrix}
approaches the evolution generated by
\begin{equation}
 a^2M_\sigma^2(\tau)
 =
 \frac{\kc}{-\tau}
\end{equation}
over the same finite interval.  The ideal Coulomb transfer function is
obtained only when the onset and termination lie outside the range that
affects the modes under consideration.

Finally, the delta-function approximation assumes
\begin{equation}
 \omega_k(\tau_j)\Delta\tau_j\ll1
\end{equation}
for each physical pulse width \(\Delta\tau_j\).  A finite width suppresses
the highest-frequency ringing and generally removes exact destructive
zeros, but it does not change the transfer-matrix organization.

\section{Deriving the universal Coulomb transfer function from the log-spaced pulse limit}
\label{sec:coulomb-transfer-derivation}

In this Appendix we derive the transfer function used in the main text directly from the exact solutions of the continuum limit of the log-spaced pulse tower.  

\paragraph{From the pulse tower to the Coulomb equation}

The dense logarithmic tower considered in the main text produces, after coarse-graining, the effective mode equation
\begin{equation}
 u_k''+
 \left[
 k^2-\frac{2}{\tau^2}+\frac{\kc}{-\tau}
 \right]u_k=0,
 \qquad \tau<0,
 \qquad \kc>0 .
 \label{eq:uk-tau-coulomb-section}
\end{equation}
Here $u_k=a\,\delta\sigma_k$ is the canonically normalized spectator fluctuation.  We introduce the dimensionless variable
\begin{equation}
 x\equiv -k\tau>0 .
\end{equation}
Since $d/d\tau=-k\,d/dx$, Eq.~\eqref{eq:uk-tau-coulomb-section} becomes
\begin{equation}
 \frac{d^2 u_k}{dx^2}+
 \left[
 1-\frac{2}{x^2}+\frac{\kc}{kx}
 \right]u_k=0 .
 \label{eq:coulomb-x-section}
\end{equation}
The standard Coulomb equation is
\begin{equation}
 \frac{d^2 u}{dx^2}+
 \left[
 1-\frac{2\eta}{x}-\frac{\ell(\ell+1)}{x^2}
 \right]u=0 .
 \label{eq:standard-coulomb-section}
\end{equation}
Therefore our system corresponds to
 $
 \ell=1$,
 $\eta=-{\kc}/{(2k)}$.
 The transfer function will follow from the behavior of the Coulomb solutions between the early-time region $x\to\infty$ and the late-time region $x\to0$.

\paragraph{Early-time normalization}

Let $F_\ell(\eta,x)$ and $G_\ell(\eta,x)$ be the regular and irregular Coulomb functions.  We use the outgoing Coulomb combination
\begin{equation}
 H^{(+)}_\ell(\eta,x)
 \equiv G_\ell(\eta,x)+iF_\ell(\eta,x) .
 \label{eq:Hplus-def-section}
\end{equation}
At large $x$ it has the unit-amplitude oscillatory form
\begin{equation}
 H^{(+)}_\ell(\eta,x)
 \sim
 \exp\left\{
 i\left[x-\eta\ln(2x)-\frac{\ell\pi}{2}+\sigma_\ell(\eta)\right]
 \right\},
 \qquad x\to\infty,
 \label{eq:Hplus-asymp-section}
\end{equation}
where
\begin{equation}
 \sigma_\ell(\eta)=\arg\Gamma(\ell+1+i\eta)
\end{equation}
is the Coulomb phase shift.  The logarithmic Coulomb phase in Eq.~\eqref{eq:Hplus-asymp-section} is a phase effect.  It does not affect the power spectrum.  Thus the properly normalized early-time solution can be written, up to an irrelevant phase, as
\begin{equation}
 u_k(x)=\frac{1}{\sqrt{2k}}H^{(+)}_1(\eta,x),
 \qquad
 \eta=-\frac{\kc}{2k} .
 \label{eq:early-solution-section}
\end{equation}
For $\kc=0$, namely $\eta=0$, this reduces to the usual de Sitter positive-frequency solution, again up to an irrelevant constant phase.

\paragraph{Late-time behavior and the coefficient of the frozen mode}

The late-time limit of the spectator fluctuation is obtained from $x\to0$. In general,
in the small-$x$ limit we have
\begin{eqnarray}
 F_\ell(\eta,x)&=&C_\ell(\eta)x^{\ell+1}+\mathcal O(x^{\ell+2})\,
 \\
  G_\ell(\eta,x)&=&\frac{1}{(2\ell+1)C_\ell(\eta)}\frac{1}{x^\ell}+\mathcal O(x^{-\ell+1}) .
\end{eqnarray}
with
\begin{equation}
 C_\ell(\eta)=
 \frac{2^\ell e^{-\pi\eta/2}|\Gamma(\ell+1+i\eta)|}{(2\ell+1)!} .
 \label{eq:Cell-section}
\end{equation}
Specializing to  $\ell=1$, the regular and irregular Coulomb functions behave as
\begin{equation}
 F_1(\eta,x)=C_1(\eta)x^2+\mathcal O(x^3),
 \label{eq:F1-small-section}
\end{equation}
while
\begin{equation}
 G_1(\eta,x)=\frac{1}{3C_1(\eta)}\frac{1}{x}+\mathcal O(1) .
 \label{eq:G1-small-section}
\end{equation}
Therefore
\begin{equation}
 H^{(+)}_1(\eta,x)
 =
 \frac{1}{3C_1(\eta)}\frac{1}{x}
 +\mathcal O(1)
 +iC_1(\eta)x^2+\cdots .
 \label{eq:Hplus-small-section}
\end{equation}
Only the $x^{-1}$ term contributes to the frozen late-time spectator fluctuation.  Indeed, in exact de Sitter,
\begin{equation}
 a(\tau)=-\frac{1}{H\tau}=\frac{k}{Hx},
\end{equation}
so that
\begin{equation}
 \delta\sigma_k=\frac{u_k}{a}=\frac{Hx}{k}u_k .
\end{equation}
Using Eq.~\eqref{eq:early-solution-section} and Eq.~\eqref{eq:Hplus-small-section}, we find
\begin{equation}
 \delta\sigma_k(\tau\to0^-)
 =
 \frac{H}{\sqrt{2k^3}}
 \frac{1}{3C_1(\eta)}
 \times e^{i\varphi(\eta)} ,
 \label{eq:sigma-late-section}
\end{equation}
where $e^{i\varphi(\eta)}$ denotes an irrelevant phase.  For $\eta=0$, one has $C_1(0)=1/3$, and Eq.~\eqref{eq:sigma-late-section} reduces to the standard massless spectator result
\begin{equation}
 \delta\sigma_k^{(0)}(\tau\to0^-)
 =
 \frac{H}{\sqrt{2k^3}}
 \times e^{i\varphi_0} .
\end{equation}
Thus the entire effect of the log-spaced pulse tower on the late-time power spectrum is encoded in the factor
\begin{equation}
 {
 \mathcal T_\sigma(k)=\left|\frac{1}{3C_1(\eta)}\right|^2,
 \qquad
 \eta=-\frac{\kc}{2k} .
 }
 \label{eq:T-C1-section}
\end{equation}
This is the first useful form of the universal transfer function.

\subsection{Evaluating the Coulomb normalization constant}

For $\ell=1$ Eq.~\eqref{eq:Cell-section} gives
\begin{equation}
 C_1(\eta)=
 \frac{1}{3}e^{-\pi\eta/2}|\Gamma(2+i\eta)| .
 \label{eq:C1-section}
\end{equation}
Substituting into Eq.~\eqref{eq:T-C1-section},
\begin{equation}
 \mathcal T_\sigma(k)
 =
 \frac{e^{\pi\eta}}{|\Gamma(2+i\eta)|^2} .
 \label{eq:T-gamma-section}
\end{equation}
We now use
\begin{equation}
 |\Gamma(2+i\eta)|^2
 =
 (1+\eta^2)|\Gamma(1+i\eta)|^2,
\end{equation}
and the standard identity
\begin{equation}
 |\Gamma(1+i\eta)|^2
 =
 \frac{\pi\eta}{\sinh(\pi\eta)} .
\end{equation}
Therefore
\begin{equation}
 {
 \mathcal T_\sigma(k)
 =
 \frac{e^{\pi\eta}\sinh(\pi\eta)}
 {\pi\eta(1+\eta^2)} ,
 \qquad
 \eta=-\frac{\kc}{2k} .
 }
 \label{eq:T-eta-section}
\end{equation}
This expression is positive also for negative $\eta$, because both $\eta$ and $\sinh(\pi\eta)$ change sign.

It is often clearer to define
\begin{equation}
 a\equiv -\eta=\frac{\kc}{2k}>0 .
\end{equation}
Then Eq.~\eqref{eq:T-eta-section} becomes
\begin{equation}
 {
 \mathcal T_\sigma(k)
 =
 \frac{1-e^{-2\pi a}}
 {2\pi a(1+a^2)} ,
 \qquad
 a=\frac{\kc}{2k} .
 }
 \label{eq:T-a-section}
\end{equation}
Equivalently,
\begin{equation}
 {
 \mathcal T_\sigma(k)
 =
 \frac{1-e^{-\pi \kc/k}}
 {\displaystyle \frac{\pi \kc}{k}
 \left(1+\frac{\kc^2}{4k^2}\right)} .
 }
 \label{eq:T-final-section}
\end{equation}
This is the universal transfer function used in the main text.


  {\small
\addcontentsline{toc}{section}{References}
\bibliographystyle{utphys}

\bibliographystyle{utcaps}
\bibliographystyle{kp}


\providecommand{\href}[2]{#2}\begingroup\raggedright\endgroup

}

\end{document}